\documentclass[journal]{IEEEtran}
\usepackage{hyperref}
\usepackage{color}
\usepackage{amsmath}
\usepackage{amssymb}
\usepackage{graphicx}
\usepackage{amsthm}
\usepackage{cite}
\usepackage[ruled,linesnumbered]{algorithm2e}
\usepackage{multirow}
\usepackage{diagbox}
\usepackage{subfigure}
\usepackage{hyperref}
\usepackage[figure]{hypcap}
\usepackage{multicol}
\usepackage{cuted} 
\usepackage{balance}  
\usepackage[table]{xcolor}    
\usepackage{booktabs}    
\usepackage{tabularx}
\usepackage{makecell}
\usepackage{array}            
\hypersetup{
  colorlinks=true,
  linkcolor=blue,
  citecolor=blue,
  urlcolor=blue
}
\usepackage{cleveref}
\usepackage{upgreek}
\usepackage{threeparttable} 
\usepackage{pdfpages}

\newcolumntype{Y}{>{\centering\arraybackslash}X}

\begin{document}
%
\title{A Data-Driven Adaptive
Impedance Matching Method Robust to Parasitic Effects}
\author{Wendong Cheng, Li Chen,~\IEEEmembership{Senior Member,~IEEE} and Weidong Wang
\thanks{This work was supported by the Natural Science Foundation of China (Grant No. 62522126) and Anhui Provincial Natural Science Foundation (No. 2308085J24). \emph{(Corresponding author: Li Chen.)}}
\thanks{Wendong Cheng, Li Chen and Weidong Wang are with the CAS Key Laboratory of Wireless Optical Communication, University of Science and Technology of China (USTC), Hefei 230027, China (e-mail: cwd01@mail.ustc.edu.cn; chenli87@ustc.edu.cn; wdwang@ustc.edu.cn).}
}

\makeatletter
\def\ps@IEEEtitlepagestyle{
  \def\@oddfoot{\mycopyrightnotice}
  \def\@evenfoot{}
}
\def\mycopyrightnotice{
  {\scriptsize
  \begin{minipage}{\textwidth}
  \centering
  
\copyright~2025 IEEE. Personal use is permitted, but republication/redistribution requires IEEE permission.
See https://www.ieee.org/publications/rights/index.html for more information.
  \end{minipage}
  }
}

\maketitle

\begin{abstract}
Adaptive impedance matching between antennas and radio frequency front-end (RFFE) power modules is essential for mobile communication systems. To address the matching performance degradation caused by parasitic effects in practical tunable matching networks (TMNs), this paper proposes a data-driven adaptive impedance matching method that avoids physical adjustment. First, we propose the residual enhanced circuit behavior modeling network (RECBM-Net), a deep learning model that maps TMN operating states to their scattering parameters (S-parameters). Then, we formulate the matching process based on the trained surrogate model as a mathematical optimization problem. We employ two classic numerical methods with different online computational overhead, namely simulated annealing particle swarm optimization (SAPSO) and adaptive moment estimation with automatic differentiation (AD-Adam), to search for the matching solution. To further reduce the online inference overhead caused by repeated forward propagation through RECBM-Net, we train an inverse mapping solver network (IMS-Net) to directly predict the optimal solution. Simulation results show that RECBM-Net accurately predicts S-parameters, achieving a mean absolute error of \(6.98 \times 10^{-5}\). Across 9000 mismatched scenarios, the compliance rate after tuning increases from 0.97\% with the analytical solution of the ideal L-network to 95.92\% with SAPSO, 93.42\% with AD-Adam, and 95\% with IMS-Net. While AD-Adam significantly reduces computational overhead, lowering the average number of RECBM-Net inferences from 2097 with SAPSO to 285, it sacrifices some accuracy. IMS-Net requires only a single inference to obtain the matching solution, resulting in minimal online overhead while maintaining excellent matching accuracy.

\end{abstract}


\IEEEpeerreviewmaketitle

\section{Introduction}\label{section_1}
\IEEEPARstart{T}{he} impedance mismatch between the antenna and the radio frequency (RF) front-end power module reduces the power transmitted to the antenna and degrades overall system performance in mobile communication systems. Furthermore, impedance mismatch adversely affects the linearity of a digital predistortion power amplifier \cite{zenteno2015output}, resulting in signal modulation distortion and spectrum expansion. More critically, the reflected power caused by the mismatch may damage the RF front-end (RFFE) components \cite{van2007power}. Therefore, achieving impedance matching is crucial for mobile devices to maintain reliable and high-performance communication.

The antenna impedance of mobile devices is influenced by multiple factors, e.g. operating frequency \cite{alibakhshikenari2019automated}, various electromagnetic interactions arising from user holding methods \cite{boyle2003performance, ogawa2001analysis}, user proximity \cite{boyle2007analysis, adams2023miniaturized}, and even user age and clothing \cite{sacco2021antenna}. The dynamic operating conditions of these devices induce impedance mismatches in the time-frequency domain. To mitigate the detrimental effects of impedance mismatch, extensive research has been conducted on adaptive impedance matching techniques for mobile communication systems.

Conventionally, adaptive impedance matching is primarily achieved through numerical methods. The numerical methods use the feedback signal reflecting the degree of mismatch to determine the optimal matching solution through extensive trial and error. The works in \cite{kong2019adaptive, van2007rf, ida2004adaptive} adopted sequential search strategies based on matching performance feedback. To enhance search efficiency, the authors in \cite{de2004rf} employed various gradient-free optimization methods to minimize the magnitude of the reflection coefficient, such as the single-step and powell algorithms. Additionally, the work in \cite{sun1999antenna, smith2013improved, ma2015automatic} employed heuristic algorithms, while the work in \cite{ogawa2003automatic} utilized gradient information to perform steepest gradient descent, reducing the number of trial-and-error iterations. The authors in \cite{xiong2019novel} observed that when the parallel inductance of the T-network is fixed, the real part of the input impedance follows a linear fractional relationship with the reactance of the first series element, and proposed a binary search algorithm to reduce the number of iterations. The disadvantages of numerical methods are their inefficiency and riskiness. As the matching network components must be physically adjusted during each trial, this tuning process introduces both amplitude and phase modulation to the radiated signal, which can lead to data corruption if tuning occurs during transmission \cite{sjoblom2005adaptive}. Therefore, numerical methods prove impractical for achieving adaptive impedance matching in mobile devices.


Compared to numerical methods, analytical methods derive the matching solution directly from the circuit model and accurate complex impedance information, thereby avoiding the iterative trial-and-error procedures. The authors in \cite{ali2013dynamic} proposed a method for real-time measurement of the complex impedance at any location within the RFFE of a handset based on perturbation theory, providing critical support for the analytical computation of the matching solution. The work of \cite{thompson2004determination} presented explicit mathematical expressions for the values of components when the ideal \(\pi\)-network is perfectly matched. Furthermore, the authors in \cite{van2009adaptive} developed an orthogonal detection technique that directly sets the real and imaginary parts of the impedance. To address imperfect matching caused by the limited tuning range of practical \(\pi\)-network, the work of \cite{gu2011analytical} proposed an analytical method to compute the optimal solution within the tunable range. Additionally, the work of \cite{gu2012new} considered the parasitic effects introduced by microelectromechanical systems (MEMS) components in the practical \(\pi\)-network and provided an analytical solution by solving the conjugate matching equation of the equivalent circuit, which remains an approximation of the practical circuit. As analytical methods are inherently model-dependent, the accuracy of the matching is determined by the degree of conformity between the practical circuit model and the model assumed by the algorithm.

Recently, powerful artificial intelligence (AI) technologies have been deployed to realize efficient and accurate adaptive impedance matching. To address impedance mismatch caused by frequency changes, the work in \cite{kim2021antenna} proposed a deep learning model that maps a target frequency response of the reflection coefficient to the corresponding L-network configuration for a given planar inverted-F antenna. Considering the complexity of the model, the authors in \cite{hasan2023adaptive} proposed a low-complexity shallow learning model based on ridge regression to achieve the same mapping. In addition, to address the impedance mismatch caused by variations in the distance between the transmitting and receiving coils in wireless power transfer systems, the work in \cite{jeong2019real} employed a neural network to establish a mapping between the equivalent load impedance and the corresponding matching solution at a specific frequency. Furthermore, considering the time-frequency domain mismatch, the work of \cite{cheng2024time} proposed a deep neural network (DNN) to directly determine the ideal \(\pi\)-network matching solution based on incomplete impedance information. The construction of large and accurate supervised datasets is critical to the success of AI-based adaptive impedance matching methods.

The above analytical and AI-based methods are applicable only to tunable matching networks (TMNs) that either neglect parasitic effects or consider only simplified cases. However, parasitic effects in practical TMNs are unavoidable and complex due to inherent limitations in the process of manufacturing and packaging of RF circuits. For example, common structures such as traces, bonding wires \cite{zhao2024novel}, vias \cite{fang2023closed}, pins \cite{xiao2023simulation}, solder balls \cite{song2022modeling}, etc., will introduce parasitic capacitance, inductance, and resistance. Particularly in mobile communication systems, TMNs are often implemented using discrete electronically switchable inductor and capacitor banks \cite{afsari2024electronically} to satisfy miniaturization constraints. The dense integration of switches, transmission lines, and non-ideal discrete components significantly aggravates parasitic effects. 

Parasitic effects render the equivalent circuit topology of a practical TMN unknown. When these effects are taken into account, matching solutions derived from analytical methods based on idealized or specific topologies often fail to meet the actual matching requirements. Similarly, for the aforementioned AI-based methods, datasets constructed from idealized topologies fail to reflect actual matching behavior, while those based on practical topologies suffer from the difficulty of acquiring accurate labels. These issues lead to degraded matching performance in AI-based methods as well. To the best of our knowledge, achieving efficient and accurate adaptive impedance matching remains an open problem when parasitic effects cause the practical TMN to deviate from its idealized topology.

To address this challenge, in this paper, we propose a data-driven adaptive impedance matching method. First, we analyze the impact of parasitic effects on the behavior of a TMN using multiport network theory. Considering the high nonlinearity and uncertainty introduced by these parasitic effects, we propose a deep learning model, residual enhanced circuit behavior modeling network (RECBM-Net), to accurately learn the system behavior. Based on the trained RECBM-Net, we further develop three matching solution determination strategies with varying online computational overhead: simulated annealing particle swarm optimization (SAPSO), adaptive moment estimation with automatic differentiation (AD-Adam), and an inverse mapping solver network (IMS-Net) that directly infers the matching solution. Our main contributions are summarized as follows:
\begin{itemize}
  \item \textbf{Behavioral modeling of parasitic effects in TMNs via multiport network theory and the DNN.} First, we analyze the advantages of using scattering parameters (S-parameters) to characterize the circuit behavior of TMNs and derive a general expression for the S-parameters of an arbitrary two-port TMN. Then, we explain the challenges associated with analytically constructing this general expression and discuss the feasibility of employing deep learning for behavioral modeling. To characterize the circuit behavior of a TMN while explicitly accounting for parasitic effects, we propose RECBM-Net, a DNN that learns the mapping between the TMN's operating state and its corresponding S-parameters.
  
  \item \textbf{Reformulating data-driven adaptive impedance matching as a mathematical optimization problem.} To address the challenges of formulating and solving conjugate matching equations due to the black-box nature of RECBM-Net in characterizing circuit behavior, we reformulate the adaptive impedance matching as a mathematical optimization problem based on the S-parameters. By varying the input to RECBM-Net, the physical adjustment of tunable elements can be virtually emulated. The objective is to identify a TMN configuration whose S-parameters closely approximate the target transmission characteristics under the matched condition.
  
  \item \textbf{Efficient matching solution determination strategies based on the RECBM-Net surrogate model.} While brute-force grid search can identify the matching network configuration, it incurs substantial online computational overhead due to repeated RECBM-Net evaluations. To reduce the cost of solving the previously formulated optimization problem, we first propose SAPSO, a hybrid algorithm that integrates global exploration with rapid convergence. In addition, we leverage automatic differentiation (AD) and employ the adaptive moment estimation (Adam) to enable efficient gradient-based search for a matching solution, further reducing computational overhead. To completely eliminate the need for repeated RECBM-Net evaluations, we introduce IMS-Net, a neural solver trained to directly predict the optimal matching solution.
\end{itemize} 

The remainder of this paper is organized as follows. Section \ref{section_2} introduces an adaptive impedance matching system and formulates the matching problem while accounting for parasitic effects. In Section \ref{section_3}, we propose a deep learning model, RECBM-Net, designed to fit the circuit behavior of the TMN. Section \ref{section_4} presents three strategies for determining matching solutions based on the surrogate model of a practical TMN, each with varying computational overhead. The matching performance on a practical L-network is evaluated in Section \ref{section_5}. Finally, Section \ref{section_6} concludes the paper.

\section{System Model}\label{section_2}
In this section, we will begin by introducing the adaptive impedance matching system. Next, we will analyze the matching solution for the ideal L-network. Finally, we will examine in detail the complexities introduced by parasitic effects and formulate an adaptive impedance matching problem that accounts for parasitic effects.
\subsection{Adaptive Impedance Matching System}\label{section_2_1}
\begin{figure}
\centering
\includegraphics[scale=0.7]{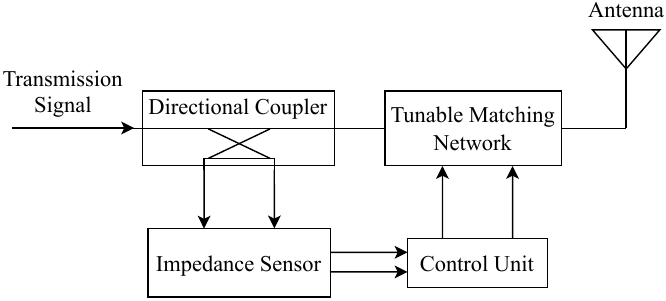}
\caption{Block diagram of an adaptive impedance matching system.}\label{system} 
\end{figure}

Consider the adaptive impedance matching system depicted in Fig. \ref{system}. It is composed of three main modules: a TMN, an impedance sensor, and a control unit. The TMN is typically composed of tunable capacitors and tunable inductors, with common configurations including the L-network and the \(\pi\)-network. It provides the desired impedance transformation by adjusting the values of the reactive components. The impedance sensor detects impedance variations by analyzing the incident and reflected signals extracted via a directional coupler. It 
can provide the control unit with key parameters such as the reflection coefficient and voltage
standing wave ratio (VSWR). The control unit employs an adaptive impedance matching method to formulate a strategy for adjusting the TMN.

The control unit generally formulates the corresponding matching strategy based on two distinct types of measurement data \cite{firrao2008automatic}. One is the data that only indicates the degree of impedance mismatch, such as reflected power and VSWR. Based on this type of data, the control unit can only adopt a trial-and-error impedance matching strategy. Impedance matching is achieved as quickly as possible by deciding the TMN configuration that is most worth trying each time. The other is the data that fully characterizes the complex impedance (e.g., reflection coefficient). Based on this type of data, the control unit knows the complex input impedance and can directly compute the matching solution based on the circuit model. In order to avoid the impracticality of frequent TMN adjustments, we focus on the latter scenario, in which the control unit can obtain the complex impedance. Also, we assume that the reflection coefficient is measured accurately.

\subsection{Impedance Matching in Ideal L-network}\label{section_2_2}
Among the various TMN configurations, the L-network is particularly appealing for its simplicity and efficiency. In the ideal L-network depicted in Fig. \ref{parasitic_effect}, the tunable capacitors \(C_{p}\) and \(C_{s}\) enable the necessary impedance transformation. The input impedance \(Z_{\text{in}}\) after impedance transformation by the L-network is given by 
\begin{equation}\label{eq:Zin}
Z_{\text{in}} = \dfrac{1}{j B_p + \dfrac{1}{Z_L + \dfrac{1}{j B_s}}},
\end{equation}
where \(B_{p}=\omega C_{p}\) and \(B_{s}=\omega C_{s}\) are the susceptances of \(C_{p}\) and \(C_{s}\), respectively, \(Z_{L}\) denotes the load impedance and \(\omega\) represents the angular frequency.

Based on Eq. (\ref{eq:Zin}), the reflection coefficient that characterizes the input impedance can be given by
\begin{equation}\label{eq:reflection_in}
\Gamma_{\rm in}=\frac{Z_{\rm in}-R_{S}}{Z_{\rm in}+R_{S}},
\end{equation}
where \(R_{S}\) represents the source impedance (typically 50 \(\Omega\)). Similarly, the load reflection coefficient for the antenna impedance \(Z_{L}\) is defined as
\begin{equation} \label{eq:reflection_load}
    \Gamma_{L}=\frac{Z_{L}-R_{S}}{Z_{L}+R_{S}}.
\end{equation}

The maximum power transfer theorem serves as the fundamental principle for tuning a TMN to achieve impedance matching for different \(\Gamma_{L}\). Maximum power is delivered when the input impedance is the complex conjugate of the source impedance. Since \(R_{S}\) is typically a purely resistive 50 \(\Omega\), this condition is reduced to \(Z_{\text{in}}=R_{S}\), which in turn implies that the reflection coefficient \(\Gamma_{\text{in}}\) measured by the impedance sensor is zero. Accordingly, the values of the \(C_p\) and \(C_s\) are adjusted to meet this requirement. Denoting the load impedance as \(Z_{L}=R_{L}+jX_{L}\) and substituting it into Eq. (\ref{eq:Zin}), we write the conjugate matching equations as
\begin{equation} \label{eq:matchcond}
\left\{
\begin{aligned}
    &\frac{B_s^2 R_L}{(B_p B_s R_L)^2 + (B_p + B_s - B_p B_s X_L)^2} = R_S, \\
    &\frac{2 B_p B_s X_L + B_s^2 X_L - B_p B_s^2 (R_L^2 + X_L^2) - B_p - B_s}
    {(B_p B_s R_L)^2 + (B_p + B_s - B_p B_s X_L)^2} = 0.
\end{aligned}
\right.
\end{equation}

By solving Eq. (\ref{eq:matchcond}), we obtain the closed-form expressions of the capacitor values required for impedance matching as
\begin{equation} \label{eq:Csolve}
\left\{
\begin{aligned}
&C_p=\frac{R_LR_S-R_L^2\pm X_L\sqrt{R_LR_S-R_L^2}}{\omega[R_LX_LR_S\pm R_LR_S\sqrt{R_LR_S-R_L^2}]}, \\
&C_s=\frac{X_L\pm\sqrt{R_LR_S-R_L^2}}{\omega[R_L^2+X_L^2-R_LR_S]}.
\end{aligned}
\right.
\end{equation}
\subsection{Impedance Matching with Parasitic Effects}\label{section_2_3}
\begin{figure}
\centering
\includegraphics[scale=0.4925]{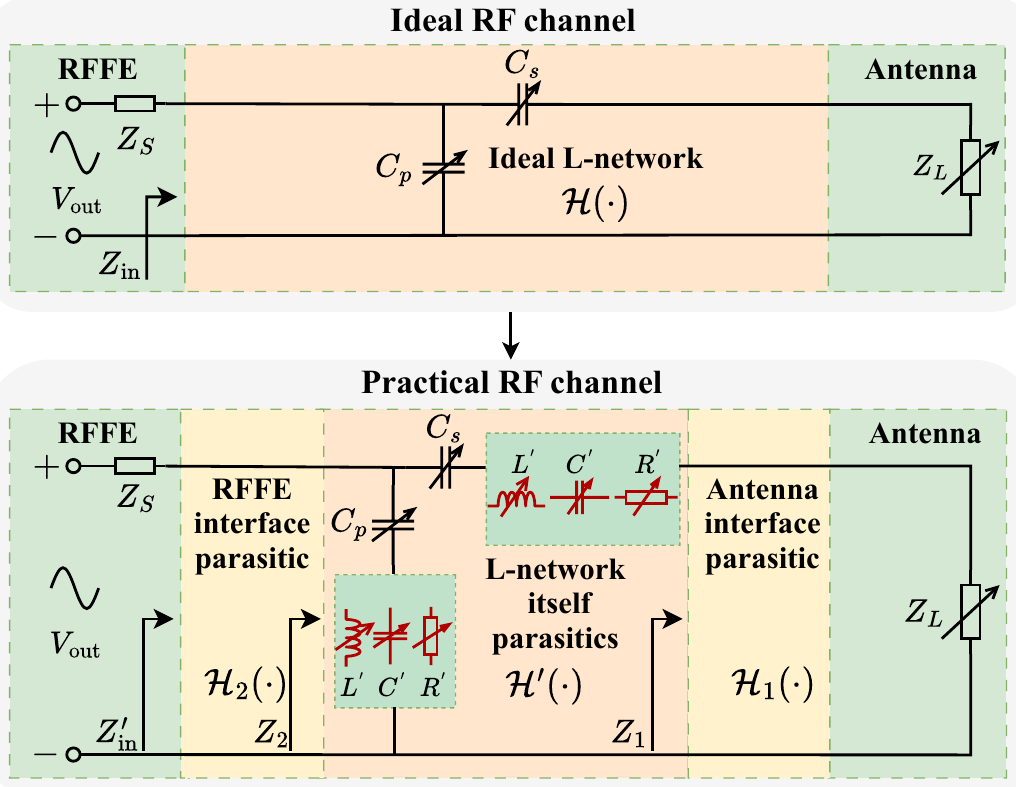}
\caption{Comparison between ideal and practical RF channels, illustrating impedance matching using an ideal L-network and the degradation caused by parasitic effects.}\label{parasitic_effect} 
\end{figure}
Considering the inherent limitations of the manufacturing process, parasitic parameters may be introduced at any location along the RF channel connecting the antenna to the transmitter’s RFFE (e.g., radio frequency integrated circuits or power amplifier modules). As shown in Fig. \ref{parasitic_effect}, parasitic effects in the RF channel arise from three key aspects. 
\begin{itemize}
  \item \textbf{L-network parasitics}:  These parasitic effects primarily stem from the manufacturing process of the TMN itself \cite{suvarna2016transformer}. Specifically, whether employing continuously tunable components (e.g., varactors or MEMS varactors) or digitally tunable components (e.g., MEMS switches, p-type-intrinsic-n-type diodes, or complementary metal-oxide-semiconductor switches), each may introduce additional parasitic capacitance, inductance, and resistance. In the ideal case, the L-network transforms the load impedance \(Z_{L}\) into the input impedance \(Z_{\text{in}}\) via the function \(\mathcal{H}(\cdot)\), as described by Eq. (\ref{eq:Zin}). The presence of parasitic effects modifies this transformation to a new mapping \(\mathcal{H}^{\prime}(\cdot)\), which maps the intermediate impedance \(Z_{1}\) to \(Z_{2}=\mathcal{H}^{\prime}(Z_{1};f)\), where \(f\) denotes the operating frequency.
  \item \textbf{RFFE interface parasitics}: These parasitic effects arise from the physical interconnection between the TMN and the RFFE. During the printed circuit board (PCB) routing, via formation, and soldering processes, connection discontinuities or non-ideal contacts can introduce additional parasitic capacitance, inductance, and resistance. The impedance transformation resulting from these parasitic effects is denoted by \(\mathcal{H}_{2}(\cdot)\), which maps the intermediate impedance \(Z_{2}\) to the final input impedance \(Z^{'}_{\text{in}}=\mathcal{H}_{2}(Z_{2};f)\).
  \item \textbf{Antenna interface parasitics}: These parasitic effects arise from the connection process between the TMN and the antenna with mechanisms similar to those discussed for the RFFE interface. The impedance transformation resulting from this portion of the parasitic effects is denoted by \(\mathcal{H}_{1}(\cdot)\),  which maps the antenna load impedance \(Z_{L}\) to the intermediate impedance \(Z_{1}=\mathcal{H}_{1}(Z_{L};f)\).
\end{itemize} 

Above all, due to the presence of parasitic effects, the impedance matching condition for ideal L-network given in Eq. (\ref{eq:matchcond}) is modified to
\begin{equation} \label{matchfunc_paras}
    \mathcal{H}_2 \left( \mathcal{H}' \left( \mathcal{H}_1(Z_L; f); f\right);f\right) = R_{S}.
\end{equation}
Neglecting these parasitic effects will lead to impedance matching failure. Taking parasitic effects into account, the overall impedance transformation is expressed as \((\mathcal{H}_2 \circ \mathcal{H}' \circ \mathcal{H}_1)(\cdot)\), which is strongly nonlinear and inherently uncertain. The strong nonlinearity arises from the fact that parasitic elements are typically superimposed on the RF channel in a mixture of series and shunt configurations. This complex interconnection results in a high-order rational function form for the input impedance. Furthermore, the inherent uncertainty stems from the unpredictable nature of parasitic parameters. Specifically, the types, values, and positions of parasitic parameters are unknown, making the exact mathematical form of \(\mathcal{H}_{1}(\cdot)\), \(\mathcal{H}_{2}(\cdot)\) and \(\mathcal{H}^{\prime}(\cdot)\) indeterminate. Consequently, the presence of parasitic effects not only complicates the relationship between the tunable components and the input impedance but also introduces inherent uncertainty, thereby significantly increasing the difficulty of adaptive impedance matching.

In summary, achieving efficient and accurate adaptive impedance matching under parasitic effects requires precise characterization of the modified system behavior \((\mathcal{H}_2 \circ \mathcal{H}' \circ \mathcal{H}_1)(\cdot)\). Based on the analysis above, it is evident that both modeling the circuit behavior of the TMN accurately in the presence of parasitic effects and determining the corresponding matching solution based on the established model remain formidable challenges. 


\section{Behavioral Modeling of Circuits with Parasitic Effects}\label{section_3}
In this section, we will first employ multiport network theory to demonstrate that S-parameters can effectively characterize the circuit behavior of a TMN in the presence of parasitic effects. Subsequently, we will introduce RECBM-Net, a deep learning model designed to fit the TMN behavior. 
\subsection{Analyzing TMN Using Multiport Network Theory}\label{section_3_1}
\begin{figure}
\centering
\includegraphics[scale=0.52]  
{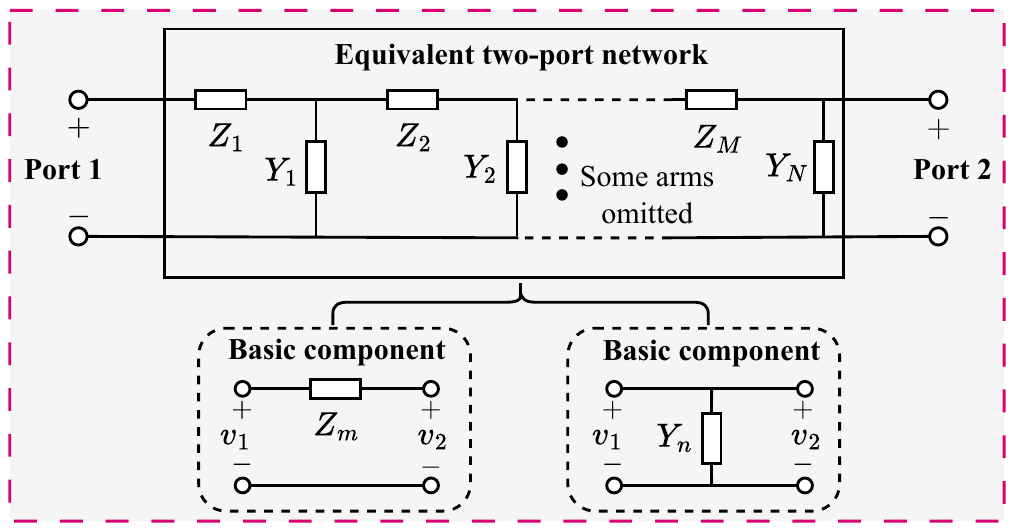}
\caption{Equivalent two-port network representation of a practical TMN, where parasitic effects result in a more complex topology consisting of \(M\) series arms and \(N\) shunt arms.}\label{Two_port_network} 
\end{figure}
Parasitic effects cause the practical TMN to deviate from the ideal L-network, resulting in a complex and uncertain topology. Therefore, it is not feasible to characterize TMN behavior from the perspective of impedance transformation. By modeling the practical TMN as a two-port network and analyzing it using multiport network theory \cite{ivrlavc2010toward}, we can explore the circuit behavior of the TMN under parasitic effects without requiring detailed knowledge of its internal structure.

Any passive two-port network representing a practical TMN can be equivalently represented as a cascade of series and shunt arms. As illustrated in Fig. \ref{Two_port_network}, the equivalent cascaded network of a practical L-network comprises \(M\) series arms and \(N\) shunt arms due to parasitic effects. The transmission (ABCD) matrix of a two-port network with a series arm is given by
\begin{equation} \label{eq:A_series}
\begin{bmatrix}
A_{\text{s},m} & B_{\text{s},m} \\
C_{\text{s},m} & D_{\text{s},m}
\end{bmatrix}=
\begin{bmatrix}
1 & Z_m \\
0 & 1
\end{bmatrix},m=1,...,M,
\end{equation}
where \(Z_{m}\) represents the impedance of the \(m\)-th series arm. Similarly, the ABCD matrix of a two-port network with a shunt arm is given by
\begin{equation} \label{eq:A_parall}
\begin{bmatrix}
A_{\text{p},n} & B_{\text{p},n} \\
C_{\text{p},n} & D_{\text{p},n}
\end{bmatrix}=
\begin{bmatrix}
1 & 0 \\
Y_{n} & 1
\end{bmatrix},n=1,...,N,
\end{equation}
where \(Y_{n}\) represents the admittance of the \(n\)-th shunt arm. By multiplying these matrices in sequence, we obtain the overall ABCD matrix of the two-port network as
\begin{equation} \label{eq:A_total}
    \begin{aligned}
        \begin{bmatrix}
            A & B \\
            C & D
        \end{bmatrix} = &
        \begin{bmatrix}
            A_{\text{s},1} & B_{\text{s},1} \\
            C_{\text{s},1} & D_{\text{s},1}
        \end{bmatrix}
        \begin{bmatrix}
            A_{\text{p},1} & B_{\text{p},1} \\
            C_{\text{p},1} & D_{\text{p},1}
        \end{bmatrix} \\ & \cdots
        \begin{bmatrix}
            A_{\text{s},M} & B_{\text{s},M} \\
            C_{\text{s},M} & D_{\text{s},M}
        \end{bmatrix}
        \begin{bmatrix}
            A_{\text{p},N} & B_{\text{p},N} \\
            C_{\text{p},N} & D_{\text{p},N}
        \end{bmatrix}.
    \end{aligned}
\end{equation}
If the first arm is shunt (rather than series), the first matrix on the right side of Eq. (\ref{eq:A_total}) is set to the identity matrix. Similarly, if the last arm is series (rather than shunt), the last matrix is set to the identity matrix. Consequently, the ABCD matrix of any passive two-port network can be expressed in the general form of Eq. (\ref{eq:A_total}), regardless of the specific arrangement of series and shunt arms. Finally, based on the conversion relationship between the ABCD matrix and the S-parameters, the S-parameters of the entire two-port network are given by
\begin{equation} \label{eq:S_total}
    \begin{bmatrix}
            S_{11} & S_{12} \\
            S_{21} & S_{22}
        \end{bmatrix} =
    \begin{bmatrix}
    \frac{A+B/Z_{0}-CZ_{0}-D}{A+B/Z_{0}+CZ_{0}+D} & \frac{2(AD-BC)}{A+B/Z_{0}+CZ_{0}+D} \\
    \frac{2}{A+B/Z_{0}+CZ_{0}+D} & \frac{-A+B/Z_{0}-CZ_{0}+D}{A+B/Z_{0}+CZ_{0}+D}
    \end{bmatrix},
\end{equation}
where \(Z_{0}\) represents the characteristic impedance. The S-parameters describe the reflection and transmission characteristics of \(n\)-port networks in high-frequency. They effectively characterize the circuit behavior of a practical TMN in the presence of parasitic effects. In addition, the S-parameters depend exclusively on the network’s internal topology and are not influenced by the load impedance.

Eqs. (\ref{eq:A_total}) and (\ref{eq:S_total}) establish the S-parameters of a practical TMN as a function of the operating frequency \(f\) and the tunable capacitors \(C_{p}\) and \(C_{s}\). We denote this mapping as
\begin{equation} \label{eq:S_behavior}
    \mathbf{S} = \mathcal{F}(f, C_{p}, C_{s}),
\end{equation}
where \(\mathcal{F}\) represents the circuit behavior under parasitic effects to be characterized. Based on the preceding network theory analysis, if the detailed topology of the practical TMN is known, a closed-form expression for the mapping \(\mathcal{F}\) can be derived. Due to parasitic effects, the exact number of series and shunt arms cannot be determined. Moreover, the specific circuit configurations of these arms are unknown, which prevents the derivation of explicit analytical expressions for \(Z_{m}\) and \(Y_{n}\). Consequently, an analytical model for the functional relationship \(\mathcal{F}\) cannot be established. 

\subsection{DNN based modeling}\label{section_3_2}
\begin{figure}
\centering
\includegraphics[scale=0.45]{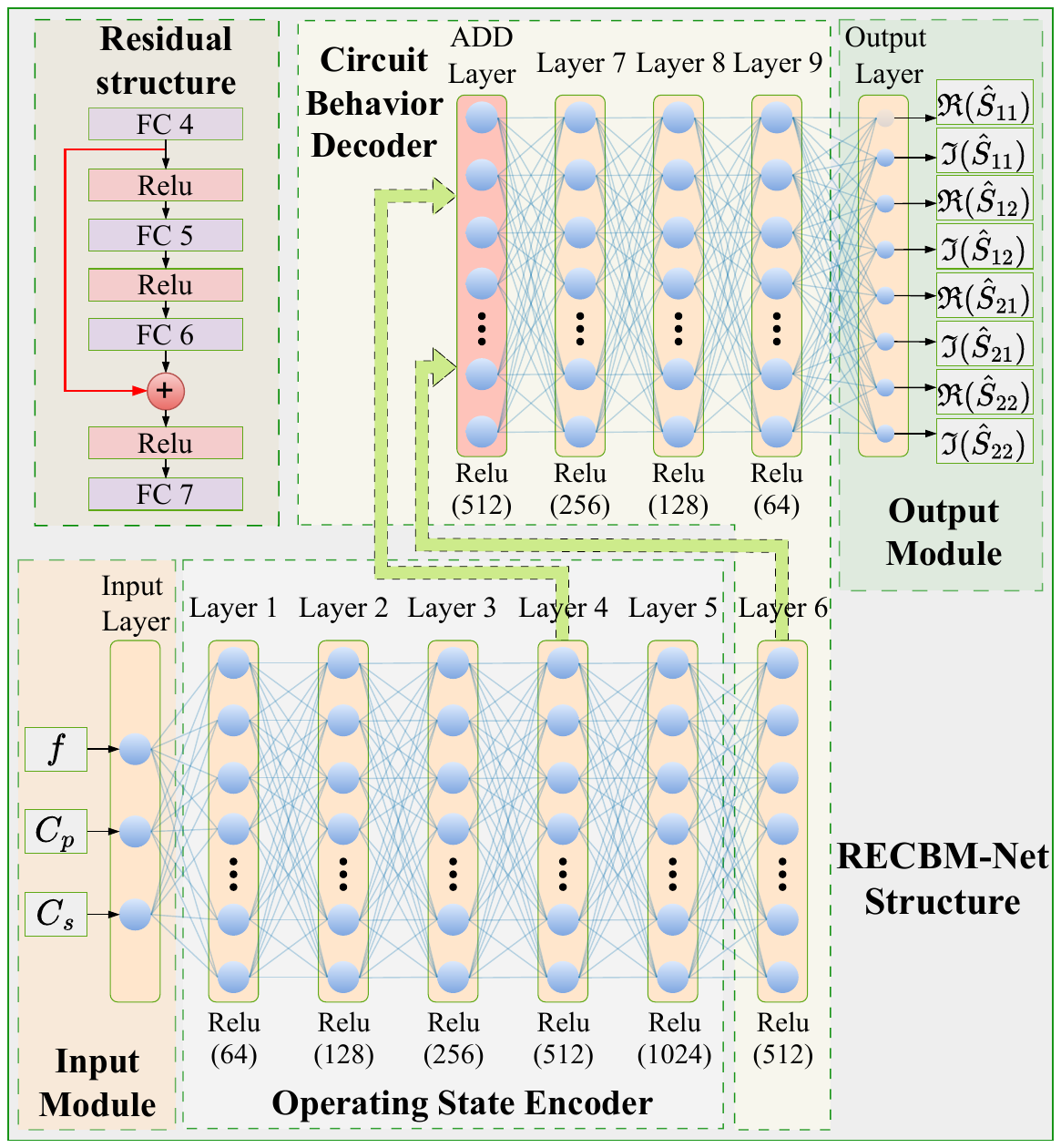}
\caption{Architecture of the proposed RECBM-Net. The network comprises five modules: an input module that receives the operating state \((f,C_{p},C_{s})\), an operating state encoder that extracts high-dimensional features, a residual structure to enhance representation learning, a circuit behavior decoder and an output module that jointly predicts the S-parameters.}\label{NN_architecture} 
\end{figure}
\(\mathcal{F}\) is an extremely complex and highly nonlinear function. According to the universal approximation theorem \cite{hornik1989multilayer}, a multilayer feedforward network with a sufficient number of hidden neurons can approximate any continuous function with arbitrary precision.  By training a DNN on the measured S-parameters of the practical TMN,  we can obtain an accurate estimate \(\hat{\mathcal{F}}\) of the mapping \(\mathcal{F}\). The preceding analysis of the fundamental mathematical representation of \(\mathcal{F}\) is crucial, as it confirms the existence of a mathematical relationship between \((f,C_{p},C_{s})\) and the S-parameters. Leveraging a data-driven approach, the DNN directly establishes an end-to-end nonlinear mapping. This method fundamentally circumvents the challenge of determining detailed topology while enabling accurate characterization of system behavior under parasitic effects.

The structure of the proposed RECBM-Net is illustrated in Fig. \ref{NN_architecture}. It consists of an input module, a TMN operating state encoder, a circuit behavior decoder, an output module and a residual structure. The input vector is defined as \(
\mathbf{x} = (f, C_{p}, C_{s})\in \mathbb{R}^{3}\). The output of the RECBM-Net is an 8-dimensional vector \(\mathbf{y} \in \mathbb{R}^{8}\), where each pair of entries corresponds to the real and imaginary components of one of the S-parameters, i.e., \(\Re(S_{11})\), \(\Im(S_{11})\), \(\Re(S_{12})\), \(\Im(S_{12})\), \(\Re(S_{21})\), \(\Im(S_{21})\), \(\Re(S_{22})\), and \(\Im(S_{22})\). 
\(\Re(\cdot)\) and \(\Im(\cdot)\) denote the real and imaginary parts, respectively. The proposed RECBM-Net employs an encoder--decoder structure with nine hidden layers. In the operating state encoder module, the hidden layer dimensionality is initially increased to capture deeper and more abstract features from the input. Specifically, the network progressively expands the feature space with 64, 128, 256, 512, and 1024 neurons. Through the operating state encoder, the 3-dimensional TMN operating state \(( f, C_{p}, C_{s})\) is transformed into a 1024-dimensional representation, thereby enabling RECBM-Net to capture the intricate nonlinear relationships between the operating state and the corresponding S-parameters. Subsequently, in the circuit behavior decoder module, the network begins to reduce the dimensionality in the later layers with 512, 256, 128, and 64 neurons, effectively decoding the learned features into a compact representation. The final output layer consists of 8 neurons, directly predicting the real and imaginary parts of the S-parameters.

All layers are fully connected (FC), with rectified linear unit (ReLU) activations applied to the hidden layers to introduce nonlinearity and mitigate the vanishing gradient problem. Additionally, the residual structure is introduced through an add layer after the sixth layer. The output of the fourth layer is directly added to that of the sixth layer, bypassing both the ReLU activations and the intermediate layers between them. Then, the summed output is passed through the ReLU activation function and continues to propagate through the subsequent layers. The residual structure improves gradient flow and alleviates network degradation while enhancing the model's ability for deeper representation learning.

\section{Matching Solution Determination Strategy Based on Surrogate Model}\label{section_4}

In this section, we first formulate the task of determining the matching solution based on a circuit behavior surrogate model as a mathematical optimization problem. Given the difficulty of solving this problem analytically, we will propose three optimization strategies with varying computational overhead.
\subsection{Formulating the Matching Optimization Problem}\label{section_4_1}
\begin{figure*}[htbp]
  \centering
  \includegraphics[scale=0.45]{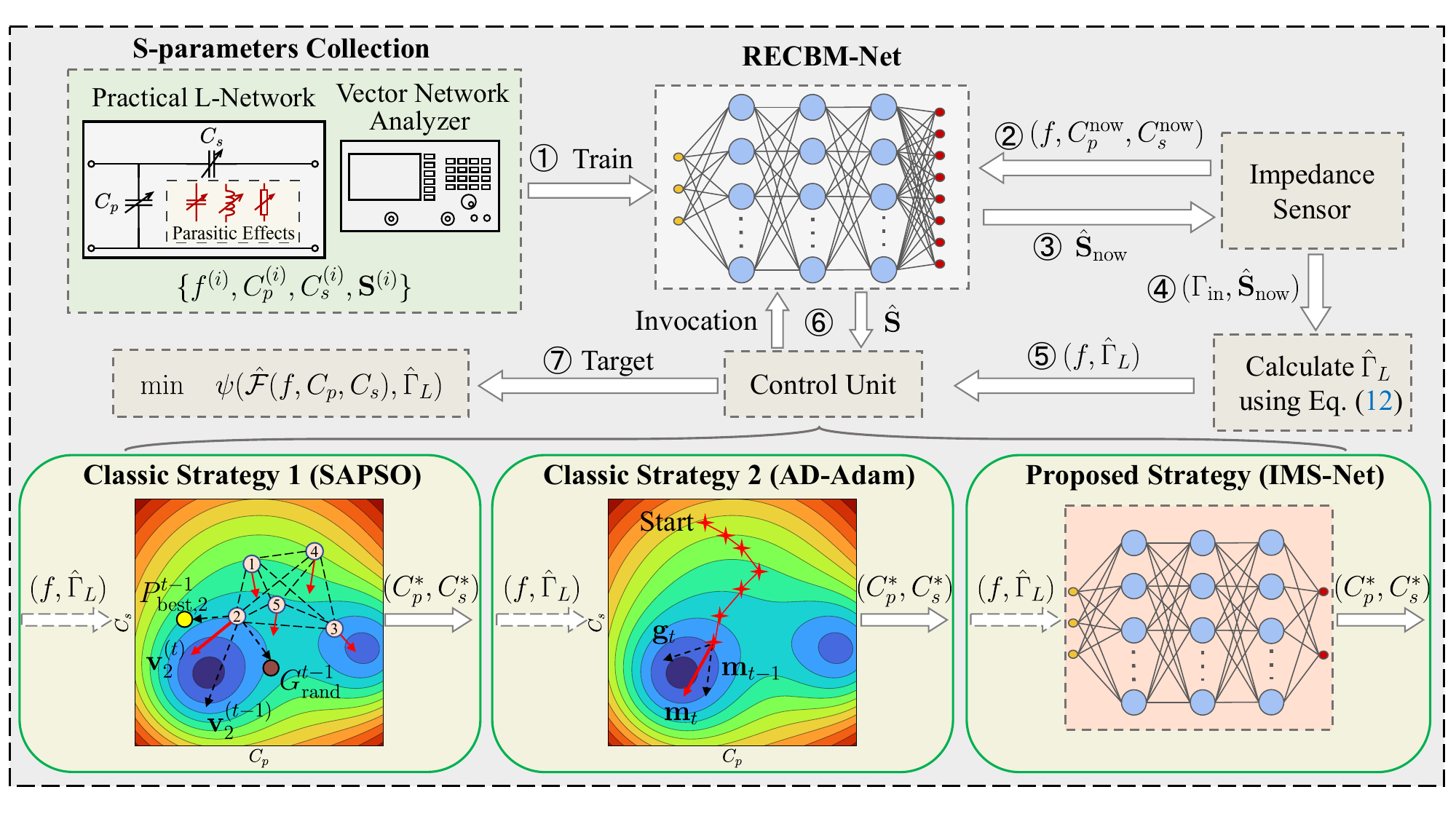}
  \caption{Overview of the proposed data-driven adaptive impedance matching method. The RECBM-Net is first trained using measured S-parameters of the practical L-network. During operation, it predicts the \(\hat{\mathbf{S}}_{\text{now}}\) corresponding to the current TMN operating state. According to Eq. (\ref{eq:tao_L}), the impedance sensor then predicts the \(\hat{\Gamma}_{L}\). Finally, the control unit determines the matching solution based on \(f\) and \(\hat{\Gamma}_{L}\). To achieve this, we propose three strategies that differ in online computational overhead.}
  \label{fig:overview}
\end{figure*}
The trained RECBM-Net accurately characterizes the behavior of the practical TMN. By employing it as the surrogate model, the matching solution can be computed entirely through simulation, eliminating the need for physical TMN adjustment. An overview of the proposed data-driven adaptive impedance matching process is shown in Fig.~\ref{fig:overview}.

The impedance sensor continuously monitors the input reflection coefficient \(\Gamma_{\text{in}}\). When an impedance mismatch is detected, the load reflection coefficient of the mismatched antenna is initially computed through the RECBM-Net. Based on the current TMN operating state \((f,C_{p}^{\text{now}},C_{s}^{\text{now}})\), the RECBM-Net predicts the corresponding S-parameters \(\hat{\mathbf{S}}_{\text{now}}\). Then, the mismatched load reflection coefficient \(\hat{\Gamma}_{L}\) is calculated by 
\begin{equation} \label{eq:tao_L}
\hat{\Gamma}_L =
\frac{\Gamma_{\text{in}} - \hat{S}_{11}}
     {\hat{S}_{12} \cdot \hat{S}_{21} + (\Gamma_{\text{in}} - \hat{S}_{11}) \cdot \hat{S}_{22}},
\end{equation}
where \(\hat{S}_{11}\), \(\hat{S}_{12}\), \(\hat{S}_{21}\) and \(\hat{S}_{22}\) are the elements of the predicted \(\hat{\mathbf{S}}_{\text{now}}\). Since the RECBM-Net can predict the S-parameters for any operating state, the input reflection coefficient \(\hat{\Gamma}_{\text{in}}\) for any configuration can be predicted by
\begin{equation} \label{eq:tao_in}
\hat{\Gamma}_{\text{in}} =
\frac{\hat{S}_{12} \cdot \hat{S}_{21} \cdot \hat{\Gamma}_L}
     {1 - \hat{\Gamma}_L \cdot \hat{S}_{22}}
+ \hat{S}_{11},
\end{equation}
where \(\hat{S}_{11}\), \(\hat{S}_{12}\), \(\hat{S}_{21}\) and \(\hat{S}_{22}\) are the elements of the predicted S-parameters \(\mathbf{\hat{S}}\) for the corresponding operating state. Thus, the matching process reduces to finding the optimal TMN configuration \((C_{p}^{*}, C_{s}^{*})\) that minimizes the predicted magnitude of the input reflection coefficient, denoted as \(|\hat{\Gamma}_{\text{in}}|\).

We denote the mapping from the predicted \(\hat{\mathbf{S}}\) and \(\hat{\Gamma}_{L}\) to the magnitude of predicted input reflection coefficient as 
\begin{equation} \label{eq:mag_taoin}
    |\hat{\Gamma}_{\text{in}}|=\psi(\hat{\mathbf{S}}, \hat{\Gamma}_{L}),
\end{equation}
where \(\psi(\cdot)\) denotes taking the magnitude of the reflection coefficient given in Eq. (\ref{eq:tao_in}). Additionally, the forward propagation of RECBM-Net for predicting \(\mathbf{\hat{S}}\) is denoted as \(\hat{\mathcal{F}}(f, C_{p}, C_{s})\). Based on this black-box model of TMN behavior, the impedance matching process can be formulated as the following mathematical optimization problem:
\begin{equation} \label{eq:optimization_prob}
\begin{aligned}
\min_{C_p, C_s} \quad & \psi\bigl(\hat{\mathcal{F}}(f, C_p, C_s), \hat{\Gamma}_{L}\bigr) \\
\text{s.t.}\quad 
& C_{p,\min} \;\le\; C_p \;\le\; C_{p,\max}, \\
& C_{s,\min} \;\le\; C_s \;\le\; C_{s,\max}.
\end{aligned}
\end{equation}
where \(C_{p,\min}\) and \(C_{p,\max}\) denote the lower and upper limits of \(C_{p}\), while \(C_{s,\min}\) and \(C_{s,\max}\) denote those of \(C_{s}\). In the current mismatched scenario, the operating frequency \(f\) and the derived load reflection coefficient \(\hat{\Gamma}_{L}\) are treated as fixed.

\begin{algorithm}
\caption{SAPSO Impedance Matching} \label{SAPSO_algorithm}
\KwIn{Number of particles \(N\), individual learning factor \(\kappa_1\), social learning factor \(\kappa_2\), cooling factor \(\lambda\), maximum iterations \(\mathcal{I}_{\max}\), load reflection \(\hat{\Gamma}_{\text{L}}\), frequency \(f\), threshold \(\varepsilon\), tuning range \(\mathcal{C}=[C_{p,\min}, C_{p,\max}] \times [C_{s,\min}, C_{s,\max}]\).}
\KwOut{Optimal matching solution \(\mathbf{C}^{*}=(C_{p}^{*}, C_{s}^{*})\).}

\textbf{Initialization:} 
\(\forall n \in \{1,\dots, N\}: \mathbf{x}_{n}^{(0)} = (x_{n,1}^{(0)}, x_{n,2}^{(0)})\),
\(\mathbf{v}_{n}^{(0)} = (v_{n,1}^{(0)}, v_{n,2}^{(0)})\), \(\Gamma_{n}^{(0)} =  \psi( \hat{\mathcal{F}}(f, \mathbf{x}_n^{(0)}), \hat{\Gamma}_L)\), \(\mathbf{P}_{\text{best},n}^{(0)} = \mathbf{x}_n^{(0)}\), \(\Gamma_{\text{best},n}^{(0)} = \Gamma_{n}^{(0)}\), \(k = \arg \min_{n} ( \Gamma_{\text{best},n}^{(0)})\), \(\mathbf{G}_{\text{best}}^{(0)} = \mathbf{x}_k^{(0)}\), \(\Gamma_{\text{best}}=\Gamma_{\text{best},k}^{(0)}\), \(T = -\Gamma_{\text{best}} / \log(0.2)\), \(\kappa = \kappa_1 + \kappa_2\), \(\chi = \frac{2}{\left| 2 - \kappa + \sqrt{\kappa^2 - 4\kappa} \right|}\).

\For{\(t = 1\) \KwTo \(\mathcal{I}_{\max}\)}{
\lIf{$\Gamma_{\text{best}} < \varepsilon$}{\textbf{break}}
    
    \(\forall n \in \{1, \dots, N\}: \ p_{n}^{(t-1)} = \exp ( - \frac{(\Gamma_{\text{best},n}^{(t-1)} - \Gamma_{\text{best}})}{T} ).\)

    \(p_{\text{sum}} = \sum_{n=1}^{N} p_{n}^{(t-1)}\), \(p_{\text{bet}} = \text{rand}(0, 1)\). \\
    \(\forall n \in \{1,\cdots,N\}:p_{n}^{(t-1)} = \frac{p_{n}^{(t-1)}}{p_{\text{sum}}}\).\\
    \For{\(k = 1\) \KwTo \(N\)}{
        \(Q_k = \sum_{n=1}^{k} p_{n}^{(t-1)}\). \\
        \If{\(p_{\text{bet}} \leq Q_k\)}{
            \(\mathbf{G}_{\text{rand}}^{(t-1)} = \mathbf{x}_k^{(t-1)}\). \\
            \textbf{break}.
        }
    }
    \For{\(n = 1\) \KwTo \(N\)}{
        \(r_{1}=\text{rand}(0,1), r_{2}=\text{rand}(0,1)\). \\
        \(
        \begin{aligned}
        \mathbf{v}_{n}^{(t)} &= \chi [\mathbf{v}_n^{(t-1)} 
        + \kappa_1  r_{1} (\mathbf{P}_{\text{best},n}^{(t-1)} - \mathbf{x}_n^{(t-1)})\\
        &\quad + \kappa_2 r_{2} (\mathbf{G}_{\text{rand}}^{(t-1)} - \mathbf{x}_n^{(t-1)})].
        \end{aligned}
        \) \\
        \(\mathbf{x}_n^{(t)} = \mathbf{x}_n^{(t-1)} + \mathbf{v}_n^{(t)}\). \\
        \If{\(\mathbf{x}_n^{(t)} \notin \mathcal{C}\)}{
        \(\Gamma_n^{(t)} = \psi(\hat{\mathcal{F}}(f, \mathbf{x}_n^{(t)}), \hat{\Gamma}_L)+P(\mathbf{x}_{n}^{(t)})\).
        }
        \Else{
            \(\Gamma_n^{(t)} = \psi(\hat{\mathcal{F}}(f, \mathbf{x}_n^{(t)}), \hat{\Gamma}_L)\).
        } \
        \text{Updated:} \(\mathbf{P}_{\text{best},n}^{(t)}\), \(\Gamma_{\text{best},n}^{(t)}\),\(\mathbf{G}_{\text{best}}^{(t)}\) \text{and} \(\Gamma_{\text{best}}\). 
        } \
    \(T = T \cdot \lambda.\)
    }
\KwRet \(\mathbf{C}^{*} = (C_{p}^{*}, C_{s}^{*})= \mathbf{G}_{\text{best}}^{(t)} \).
\end{algorithm}

\begin{algorithm} 
\caption{AD-Adam Impedance Matching}\label{Ad-Adam_algorithm}
\KwIn{Initial solution \(\boldsymbol{\theta}_0\), learning rate \(\alpha\), exponential decay rates \(\beta_1\), \(\beta_2\), stability constant \(\epsilon\), maximum iterations \(\mathcal{I}_{\max}\),  load reflection coefficient \(\hat{\Gamma}_{\text{L}}\), frequency \(f\), threshold \(\varepsilon\), tuning range \(\mathcal{C}=[C_{p,\min}, C_{p,\max}] \times [C_{s,\min}, C_{s,\max}]\).}
\KwOut{Optimal matching solution \(\mathbf{C}^{*}=(C_{p}^{*},C_{s}^{*})\).}

\textbf{Initialization:}
\( t=0\), \(\boldsymbol{\theta}^{(0)}= \boldsymbol{\theta}_0\), \(\mathbf{m}_{0}=\mathbf{0}\),\(\mathbf{v}_{0}=\mathbf{0}\).\\
\For{\(t = 1\) \KwTo \(\mathcal{I}_{\max}\)}{
    Compute the gradient using AD: \(\mathbf{g}_{t}=\nabla_{\boldsymbol{\theta}} \psi(\hat{\mathcal{F}}(f,\boldsymbol{\theta}^{(t-1)}),\,\hat{\Gamma}_L)\).\\
    Update \(\mathbf{m}_t\): \(\mathbf{m}_{t}=\beta_1 \mathbf{m}_{t-1} + (1-\beta_1)\mathbf{g}_t\).\\
    Update \(\mathbf{v}_t\): \(\mathbf{v}_{t}=\beta_2 \mathbf{v}_{t-1} + (1-\beta_2)(\mathbf{g}_t \odot \mathbf{g}_t)\).\\
    Compute bias-corrected estimates:
    \(\hat{\mathbf{m}}_t=\dfrac{\mathbf{m}_t}{1-\beta_1^t}, \hat{\mathbf{v}}_t=\dfrac{\mathbf{v}_t}{1-\beta_2^t}\).\\
    Update matching solution: \(\boldsymbol{\theta}^{(t)}=\boldsymbol{\theta}^{(t-1)}-\alpha\dfrac{\hat{\mathbf{m}}_{t}}{\sqrt{\hat{\mathbf{v}}_t}+\epsilon}\).\\
    Project \(\boldsymbol{\theta}^{(t)}\) onto feasible set \(\mathcal{C}\): \(\boldsymbol{\theta}^{(t)} = \mathrm{Proj}_{\mathcal{C}}(\boldsymbol{\theta}^{(t)}) \).\\
    Evaluate reflection coefficient magnitude: \(\Gamma_{\text{in}}^{(t)}= \psi( \hat{\mathcal{F}}(f, \boldsymbol{\theta}^{(t)}), \hat{\Gamma}_L )\). \\
     \If{\(\Gamma_{\text{in}}^{(t)} < \varepsilon\)}{
        \textbf{break}.
    }
}
\KwRet \(\mathbf{C}^*=(C_{p}^{*}, C_{s}^{*})=\boldsymbol{\theta}^{(t)}\).
\end{algorithm}

\subsection{Strategies for Solving the Matching Problem}\label{section_4_2}
Given that the RECBM-Net comprises multiple hidden layers with numerous neurons and activation functions, \(\hat{\mathcal{F}}(\cdot)\) is an extremely complex nonlinear function. Embedding \(\hat{\mathcal{F}}(\cdot)\) within the objective function \(\psi(\cdot)\) further complicates the derivation of an analytical optimal solution. To address this, we first introduce two classic numerical optimization strategies: SAPSO and AD-Adam. These methods require varying numbers of evaluations of RECBM-Net to compute the reflection coefficient for each TMN configuration, leading to considerable online computational overhead. To eliminate the need for repeated RECBM-Net inference during optimization, we further propose an IMS-Net for RECBM-Net to directly obtain the matching solution.

\textbf{Classic strategy 1 (SAPSO):} SAPSO combines the fast convergence and high accuracy of particle swarm optimization (PSO) with the ability of simulated annealing (SA) to escape local optima by occasionally accepting worse solutions. This hybridization enhances the global search for an optimal TMN configuration. Using SAPSO to obtain the matching solution involves three major steps.

\begin{itemize}
  \item \textit{Initialization of the population}:  
  Let \(n\) index the particles, and \(t\) index the iterations. At iteration \(t=0\), set each particle's initial position \(\mathbf{x}_n^{(0)}\) (representing a candidate TMN configuration) and velocity \(\mathbf{v}_n^{(0)}\) within the given constraints. For each particle \(n\), compute the initial fitness \(\Gamma_{n}^{(0)}\) and record both its personal best position \(\mathbf{P}_{\text{best},n}^{(0)}\) and the global best position \(\mathbf{G}_{\text{best}}^{(0)}\) among all particles. Based on \(\mathbf{G}_{\text{best}}^{(0)}\), determine the initial temperature \(T\) for the simulated annealing component. Additionally, initialize the compression factor \(\chi\) and the cooling factor \(\lambda\).
  \item \textit{Velocity and position updates}: At iteration \(t\), first calculate the normalized Metropolis acceptance probabilities \(p_{n}^{(t-1)}\) for each particle \(n\) based on its personal best fitness \(\Gamma_{\text{best},n}^{(t-1)}\). \(\Gamma_{\text{best},n}^{(t-1)}\) denotes the best fitness value achieved by particle \(n\) up to the current iteration. Next, perform a roulette wheel selection using \(p_{n}^{(t-1)}\) to obtain a candidate global best position \(\mathbf{G}_{\text{rand}}^{(t-1)}\). Then, update each particle’s velocity \(\mathbf{v}_{n}^{(t)}\) and position \(\mathbf{x}_{n}^{(t)}\) by taking into account both \(\mathbf{P}_{\text{best},n}^{(t-1)}\) and \(\mathbf{G}_{\text{rand}}^{(t-1)}\). If the updated position \(\mathbf{x}_{n}^{(t)}\) violates the predefined bounds of the tunable capacitors, we add a penalty term \(P(\mathbf{x}_{n}^{(t)})\) to the predicted input reflection coefficient magnitude \(\psi(\hat{\mathcal{F}}(f, \mathbf{x}_n^{(t)}), \hat{\Gamma}_L)\), yielding the penalized fitness \(\Gamma_{n}^{(t)}\) to discourage infeasible solutions. Accordingly, the penalty term is implemented using the death penalty function, with the penalty constant fixed at 2000. The rationale for this selection is detailed in Appendix A of Supplementary Material. After this update, refresh the personal best positions \(\mathbf{P}_{\text{best},n}^{(t)}\) and their corresponding reflection coefficients \(\Gamma_{\text{best},n}^{(t)}\), as well as the global best position \(\mathbf{G}_{\text{best}}^{(t)}\) and its corresponding coefficient \(\Gamma_{\text{best}}\). Finally, decrease the temperature \(T\).
  \item \textit{Acceptance criteria and termination}: During each iteration, if the global best fitness \(\Gamma_{\text{best}}\) falls below a predefined threshold \(\varepsilon\), the algorithm terminates early and outputs the corresponding optimal TMN configuration \((C_{p}^{*},C_{s}^{*})\). Otherwise, the algorithm continues until the maximum number of iterations \(\mathcal{I}_{\max}\) is reached and returns the best solution found.
\end{itemize} 

SAPSO is a non-gradient numerical optimization method with the advantage of high accuracy. It updates parameters by evaluating the fitness of all particles in each iteration. Since computing each particle’s fitness requires invoking the RECBM-Net to infer the S-parameters, it is computationally intensive. Algorithm \ref{SAPSO_algorithm} provides the detailed procedure for determining the matching solution based on SAPSO.

\textbf{Classic strategy 2 (AD-Adam):} To reduce the number of RECBM-Net evaluations, we adopt Adam \cite{kingma2014adam}, a gradient-based optimization method that efficiently updates parameters using first and second moment estimates of gradients. Adam adaptively adjusts the learning rate of each parameter, thereby achieving faster convergence. The effectiveness of Adam critically depends on accurate gradient computation. AD \cite{bartholomew2000automatic} enables the exact computation of gradients for complex nonlinear functions by constructing computational graphs and systematically applying the chain rule. It overcomes the limitations of both symbolic and numerical differentiation. Therefore, AD is used to compute the gradient of the objective function \(\psi\bigl(\hat{\mathcal{F}}(f, C_p, C_s), \hat{\Gamma}_{L}\bigr)\) with respect to \(C_{p}\) and \(C_{s}\), which becomes highly nonlinear due to the embedding of RECBM-Net. When performing backpropagation to compute the gradients, adjoint variables are propagated backward through the computational graph, and the local gradient of each elementary operator is evaluated once. Consequently, the overall computational overhead of a gradient evaluation is approximately equivalent to that of an additional forward pass. The main steps to determine the optimal matching solution using AD-Adam are as follows.

\begin{itemize}
  \item \textit{Initialization}:  
  Set the learning rate \(\alpha\) and the exponential decay rates \(\beta_{1}\), \(\beta_{2}\). Initialize the matching solution as \(\boldsymbol{\theta}_{0}=(C_{p}^{(0)},C_{s}^{(0)})\). Additionally, initialize the first-order moment estimate \(\mathbf{m}_{0}\) (momentum term) and the second-order moment estimate \(\mathbf{v}_{0}\) for the initial gradient \(\mathbf{g}_{0}\).
  \item \textit{Gradient-based parameter update}: In each iteration \(t\), compute the gradient \(\mathbf{g}_{t}\) of the objective function with respect to \(\boldsymbol{\theta}^{(t-1)}\) using AD. Then, update the first-order moment estimate \(\mathbf{m}_{t}\) and second-order moment estimate \(\mathbf{v}_{t}\) based on \(\mathbf{g}_{t}\), where \(\odot\) denotes the Hadamard product. Compute the bias-corrected first-order moment estimate \(\hat{\mathbf{m}}_{t}\) and the bias-corrected second-order moment estimate \(\hat{\mathbf{v}}_{t}\). Next, update the matching solution \(\boldsymbol{\theta}^{(t)}\) based on \(\hat{\mathbf{m}}_{t}\) and \(\hat{\mathbf{v}}_{t}\). After the update, \(\boldsymbol{\theta}^{(t)}\) is projected onto the feasible set \(\mathcal{C}\) to ensure that each parameter remains within its lower and upper limits, where the projection is defined as $\mathrm{Proj}_{\mathcal{C}}(\boldsymbol{\theta}^{(t)}) = \arg\min_{\boldsymbol{\theta} \in \mathcal{C}} \| \boldsymbol{\theta} - \boldsymbol{\theta}^{(t)} \|_2$.
  \item \textit{Acceptance criteria and termination}: At each iteration, if the predicted input reflection coefficient \(\Gamma_{\text{in}}^{(t)}\) at \(\boldsymbol{\theta}^{(t)}\) is lower than the threshold \(\varepsilon\), the optimization terminates and outputs the optimal configuration \((C_{p}^{*},C_{s}^{*})\). Otherwise, it proceeds until the maximum iteration number \(\mathcal{I}_{\max}\) is reached.
\end{itemize} 

Compared with SAPSO, AD-Adam requires only a single RECBM-Net inference per iteration, thereby significantly reducing online computational overhead. But its performance may degrade on highly nonconvex objectives, making it more susceptible to local minima and less accurate than SAPSO. Algorithm \ref{Ad-Adam_algorithm} provides the detailed procedure for impedance matching using AD-Adam.

\textbf{Proposed strategy (IMS-Net):} The two numerical optimization methods discussed above necessitate repeated invocations of RECBM-Net for S-parameter inference, which incurs significant computational overhead. To reduce this burden, we propose a novel strategy that trains IMS-Net to directly predict the matching solution without iterative DNN evaluation. 

Given a predicted \(\hat{\Gamma}_L\) and \(f\) in the current mismatched scenario, the optimal impedance matching solution \((C_p^*, C_s^*)\) satisfies the implicit equation \( \psi\bigl(\hat{\mathcal{F}}(f, C_p^*, C_s^*), \hat{\Gamma}_{L}\bigr)=0\), assuming a perfect match. This solution can be obtained in two steps. First, the predicted \(\hat{\mathbf{S}}\) that satisfies the matching condition is computed via the inverse mapping \(\psi^{-1}\). In other words, we solve the equation
\begin{equation} \label{eq:solve_S}
    \frac{\hat{S}_{12} \cdot \hat{S}_{21} \cdot \hat{\Gamma}_L}
     {1 - \hat{\Gamma}_L \cdot \hat{S}_{22}}
+ \hat{S}_{11}=0.
\end{equation}
Subsequently, the inverse mapping \(\hat{\mathcal{F}}^{-1}\) is applied to retrieve the circuit configuration corresponding to the obtained scattering property \(\hat{\mathbf{S}}\). Accordingly, the matching solution can be expressed as
\begin{equation} \label{eq:solve_S_function}
    (C_{p}^{*}, C_{s}^{*}) = \mathcal{G}(f,\hat{\Gamma}_{L}),
\end{equation}
where the composite mapping is defined as \(\mathcal{G}=\hat{\mathcal{F}}^{-1}\circ\psi^{-1}\). 

Ideally, accurately modeling the composite mapping \(\mathcal{G}\) would enable the direct computation of the matching solution. It should be noted that the predicted \(\hat{\mathbf{S}}\) from RECBM-Net may not strictly satisfy typical TMN constraints such as passivity (\( \mathbf{I} - \hat{\mathbf{S}}^\dagger \hat{\mathbf{S}} \succeq 0 \)) or reciprocity (\(\hat{S}_{12} = \hat{S}_{21}\)). Consequently, relying solely on Eq. (\ref{eq:solve_S}) to determine \(\hat{\mathbf{S}}\) results in an underdetermined system with infinitely many solutions, meaning that the mapping \(\psi^{-1}\) is multi-valued. Moreover, the circuit behavior \(\hat{\mathcal{F}}\) learned by RECBM-Net does not guarantee that its inverse \(\hat{\mathcal{F}}^{-1}\) is a single-valued mapping. Above all, the composite mapping \(\mathcal{G}\) may also be multi-valued mapping. 

To address the challenges posed by the inherent multi-valued nature of \(\mathcal{G}\) and the absence of its analytical expression, we first construct a dataset that enforces a single-valued mapping, and then train an IMS-Net to accurately approximate \(\mathcal{G}\). The main process for obtaining matching solution using IMS-Net is summarized as follows.
\begin{itemize}
  \item \textit{Dataset preparation}: Within the TMN’s operating frequency band and tunable range, an exhaustive grid is constructed by traversing all possible combinations of frequencies and TMN configurations at a predetermined granularity. Each sample is denoted as \(\mathbf{x}^{(i)} = \left\{ f^{(i)}, C_p^{(i)}, C_s^{(i)} \right\}_{i=1}^{N_d}\), where \(N_{d}\) is the total number of samples. Subsequently, the corresponding \(\hat{\mathbf{S}}^{(i)}\) for each configuration \(\mathbf{x}^{(i)}\) is predicted using the trained RECBM-Net. Assuming perfect input matching, each predicted \(\hat{\mathbf{S}}^{(i)}\) and \(\Gamma_{\text{in}}=0\) are substituted into Eq. (\ref{eq:tao_L}) to compute the corresponding load reflection coefficient \(\hat{\Gamma}_{L}^{(i)}\). A supervised learning dataset is then constructed as \(\left\{ f^{(i)}, \Re (\hat{\Gamma}_{L}^{(i)}),\Im (\hat{\Gamma}_{L}^{(i)}), C_p^{(i)}, C_s^{(i)} \right\}_{i=1}^{N_d}\), where the first three dimensions serve as input features and the last two as output labels. This approach not only calculates a large number of load impedances that can be matched by the practical TMN as predicted by RECBM-Net, but also directly provides the corresponding perfect matching solutions without solving the mathematical optimization problem in Eq. (\ref{eq:optimization_prob}). Moreover, it guarantees a single-valued mapping from \((f, \hat{\Gamma}_{L})\) to the matching solution.
  
  \item \textit{Offline training}: IMS-Net employs the same architecture as RECBM-Net, with the only difference being in the output dimension (two neurons in the output layer). The dataset is partitioned into training and validation sets. IMS-Net is trained on the former to learn the mapping \(\mathcal{G}\), and its generalization performance is evaluated on the latter.

  \item \textit{Online inference}: In the deployment phase, the trained IMS-Net takes the predicted load reflection coefficient and operating frequency as input, and directly outputs the corresponding optimal matching solution.
\end{itemize} 

Compared with the two numerical methods discussed above, the integration of IMS-Net with RECBM-Net completely eliminates repeated RECBM-Net evaluations, thereby minimizing online computational overhead. The matching process requires only two inferences: one from RECBM-Net, to predict the current TMN S-parameters and compute \(\hat{\Gamma}_{L}\), and another from IMS-Net to directly infer the matching solution. Moreover, subsequent simulation results demonstrate that IMS-Net achieves exceptionally high accuracy. It is worth noting that IMS-Net functions as a dedicated inverse solver that is tightly coupled with its corresponding RECBM-Net. Therefore, any modification to the RECBM-Net parameters (e.g., learning different TMN circuit behaviors) necessitates retraining the associated IMS-Net.

\section{Numerical Results and Discussion}\label{section_5}
In this section, we will first utilize RECBM-Net to characterize the circuit behavior of a practical L-network and evaluate its fitting accuracy in detail. Then, we will simulate extensive mismatched scenarios to verify the performance of the proposed data-driven adaptive impedance matching method.

All algorithm development and testing are carried out in a Python environment (version 3.8.20). The hardware platform is a workstation equipped with an Intel Xeon Gold 5218 central processing unit (CPU) @ 2.30 GHz and four NVIDIA GeForce RTX 2080 Ti graphics processing units (GPUs). The analytical method for computing the ABCD matrix to indirectly derive the S-parameters, as described in Section \ref{section_3_1}, is implemented in Python and verified to be consistent with simulation results obtained from Keysight Advanced Design System. This implementation is employed to generate the training and testing datasets for RECBM-Net by computing S-parameters across various circuit configurations. Similarly, the training and testing datasets for IMS-Net, as well as the simulated mismatched scenarios used for performance evaluation, are all generated in Python. The detailed data generation procedures are provided in Sections \ref{section_4_2} and \ref{section_5_2}. Additionally, both RECBM-Net and IMS-Net are trained using the PyTorch framework (version 2.4.1). The SAPSO and AD-Adam are further refined through hyperparameter tuning to improve performance. Finally, the proposed matching solution determination strategies are implemented in Python and employed to validate their end-to-end matching performance.

\subsection{Accuracy of RECBM-Net in Modeling Circuit Behavior}\label{section_5_1}
\begin{figure}
\centering
\includegraphics[scale=0.418]{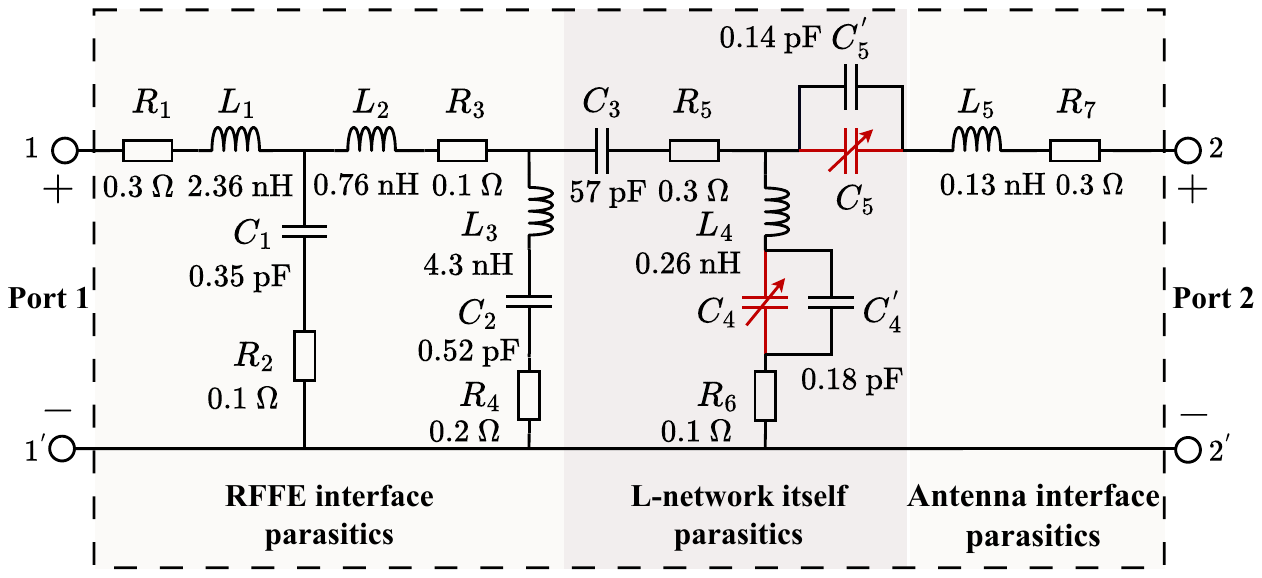}
\caption{Schematic of the L-network where 17 parasitic elements are added across the RFFE interface, the L-network itself, and the antenna interface to emulate realistic circuit behavior in practical TMNs.}\label{practical_Lnetwork} 
\end{figure}

To simulate the deviation of circuit behavior from the ideal matching network caused by practical parasitic effects, we introduce additional parasitic elements into the ideal L-network structure. Further, ladder circuits with alternating series and parallel branches are incorporated to simulate additional parasitic effects. Fig. \ref{practical_Lnetwork} illustrates the simulation of an L-network featuring 17 parasitic elements. The ladder topology is adopted because it is widely used in engineering to approximate the system response of practical devices. According to vector fitting techniques \cite{gustavsen2002rational} and network synthesis theory \cite{otomo2020synthesis}, the system response of a practical device under test (DUT) can be approximated by rational functions, which can subsequently be synthesized into a ladder-type equivalent circuit. Therefore, based on this structure, we randomly introduce a large number of additional inductors, capacitors, and resistors, with their values also assigned randomly. The objective is to construct a high-order, sufficiently complex test circuit to validate our data-driven impedance matching method. In this circuit, analytical matching solutions derived from the ideal L-network model fail completely. In our proposed method, the internal circuit structure is treated as unknown, in order to verify its effectiveness when parasitic effects obscure the practical configuration.

We generate the S-parameter dataset for the practical L-network using the Python script that analytically computes the S-parameters of a given two-port network. We systematically sweep through all combinations of operating frequency \(f\) and the values of tunable capacitors \(C_{4}\) and \(C_{5}\). Specifically, \(f\) is varied from 1.5 GHz to 2 GHz in increments of 0.02 GHz, while \(C_{4}\) and \(C_{5}\) are adjusted from 0 to 10 pF in 0.02 pF steps. This comprehensive parameter sweep enables the construction of an S-parameter dataset that accurately characterizes the practical TMN behavior, comprising a total of 6,500,000 samples. The generation of this dataset takes 202.28 seconds on an Intel Xeon Gold 5218 CPU.

\begin{figure}
\centering
\includegraphics[width=0.6\linewidth]{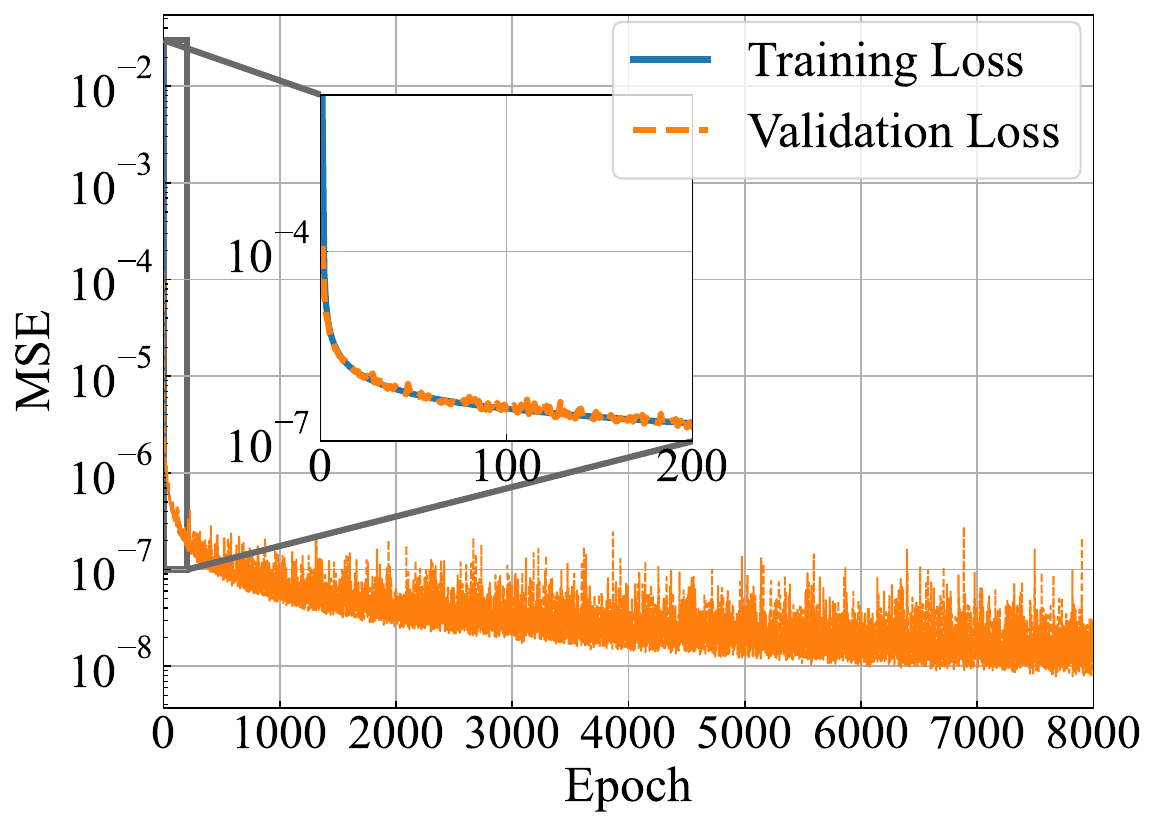}
\caption{Training and validation loss (MSE) over epochs.}\label{loss_curve} 
\end{figure}

\begin{figure*}[!t]
  \centering
  \subfigure[\(S_{11}\)]{
    \includegraphics[width=0.238\textwidth]{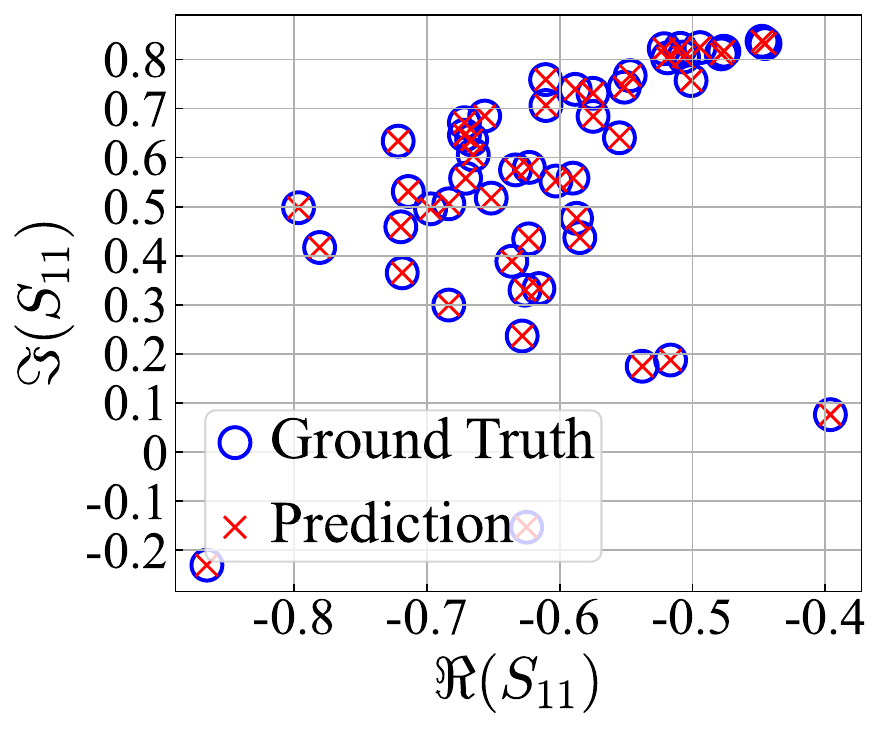}
    \label{fig:sub_a}
  } \hspace{-1.2em}
  \subfigure[\(S_{12}\)]{
    \includegraphics[width=0.238\textwidth]{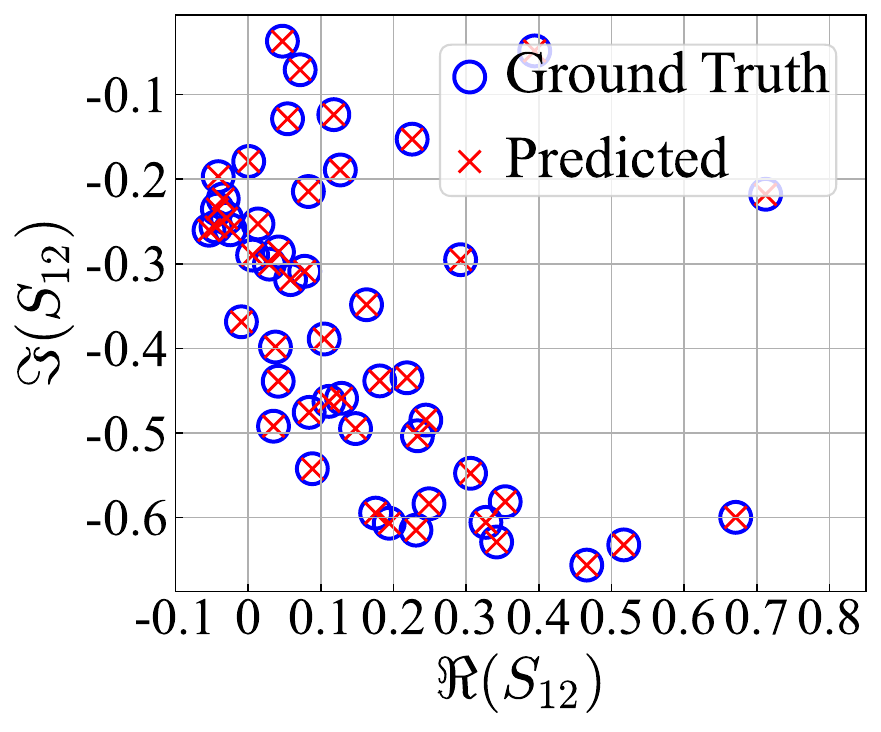}
    \label{fig:sub_b}
  } \hspace{-1.2em}
  \subfigure[\(S_{21}\)]{
    \includegraphics[width=0.238\textwidth]{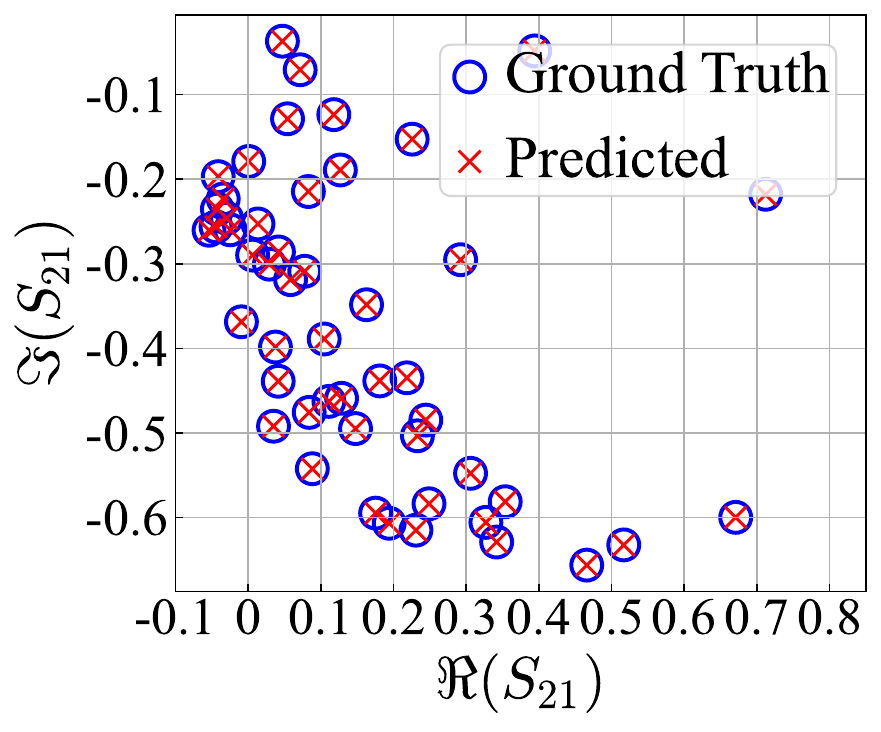}
    \label{fig:sub_c}
  } \hspace{-1.2em}
  \subfigure[\(S_{22}\)]{
    \includegraphics[width=0.238\textwidth]{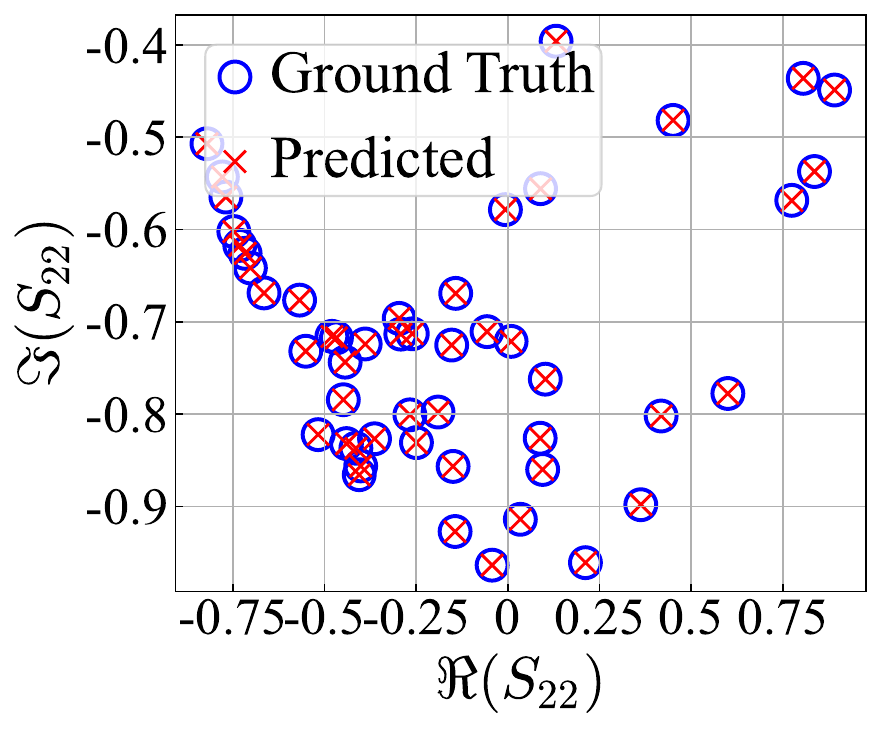}
    \label{fig:sub_d}
  }
  \caption{Comparison between predicted and ground truth S-parameters using RECBM-Net. A total of 50 samples are randomly selected from the test set containing 100,000 samples. The blue circles denote the ground truth, while the red crosses indicate the predicted values.}
  \label{true_pred}
\end{figure*}

The dataset is divided into a training set (\(80\%\)) and a validation set (\(20\%\)). The training set is used to optimize the model parameters, while the validation set serves to evaluate the model's generalization performance. Since the input features include frequency and tunable capacitor values with different numerical scales, we apply min–max normalization to all features to ensure training stability. In terms of the loss function, we employ the mean squared error (MSE) to quantify the discrepancy between the model's prediction and the actual values. It is defined as
\begin{equation} \label{loss_func}
    L(\theta) = \frac{1}{8N} \sum_{i=1}^{N} \left\| \mathbf{y}_i - \hat{\mathbf{y}}_i \right\|_2^2 ,
\end{equation}
where \(L(\theta)\) measures the average difference between the predicted vectors 
\(\hat{\mathbf{y}}_i \in \mathbb{R}^8\) and the ground truth vectors \(\mathbf{y}_i \in \mathbb{R}^8\) across all \(N\) training samples. The Adam optimizer is selected to train the RECBM-Net with an initial learning rate set to \(5 \times 10^{-8}\). A batch size of 512 is utilized, and the model is trained for 8000 epochs. Training is performed with PyTorch’s DistributedDataParallel (DDP) framework on four NVIDIA GeForce RTX 2080 Ti GPUs, requiring approximately 59.39 hours in total. 

\begin{table}
\centering
\caption{Mean Prediction Error of S-Parameters Using RECBM-Net.}
\label{tab:mean_error}
\begin{tabular}{c c c c c}
\toprule
\multirow{2}{*}{\textbf{S-Parameters}} & 
\multicolumn{2}{c}{\textbf{Mean Absolute Error}} & 
\multicolumn{2}{c}{\textbf{Mean Relative Error}} \\
\cmidrule(lr){2-3}\cmidrule(lr){4-5}
 & \textbf{Real} & \textbf{Imag} & \textbf{Real} & \textbf{Imag} \\
\midrule
\rowcolor{blue!10} \(S_{11}\) & \(7.2 \times 10^{-5}\) & \(1.0 \times 10^{-4}\) & 0.012\(\%\) & 0.084\(\%\) \\
\(S_{12}\) & \(5.3 \times 10^{-5}\) & \(5.6 \times 10^{-5}\) & 0.32\(\%\) & 0.039\(\%\) \\
\rowcolor{blue!10} \(S_{21}\) & \(5.3 \times 10^{-5}\) & \(5.6\times 10^{-5}\) & 0.319\(\%\) & 0.038\(\%\) \\
\(S_{22}\) & \(9.2 \times 10^{-5}\) & \(7.6 \times 10^{-5}\) & 0.163\(\%\) & 0.012\(\%\) \\
\bottomrule
\end{tabular}
\end{table}

\begin{figure}    
  \centering
  \subfigure[ECDF of absolute error.]{
    \includegraphics[width=0.45\linewidth]{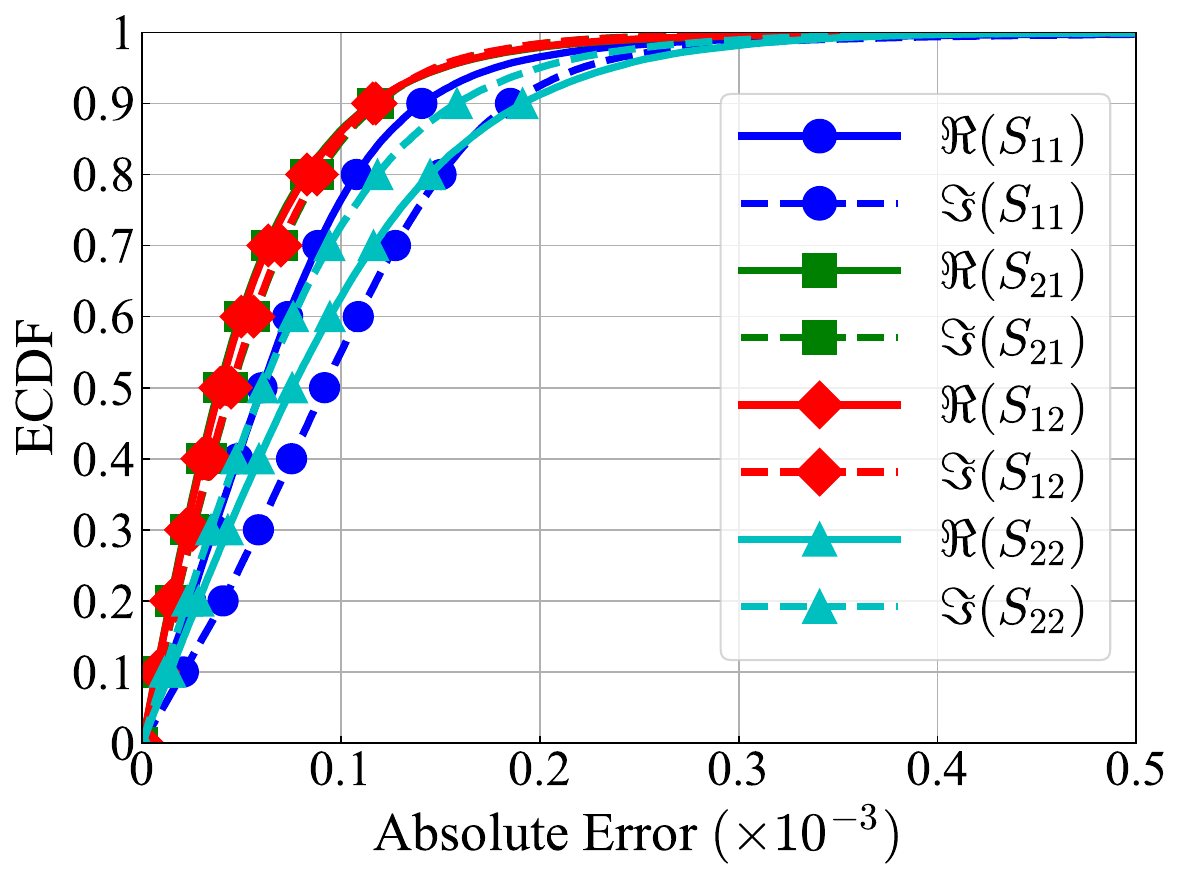}
    \label{fig:absolute_error}
  }
  \subfigure[ECDF of relative error.]{
    \includegraphics[width=0.45\linewidth]{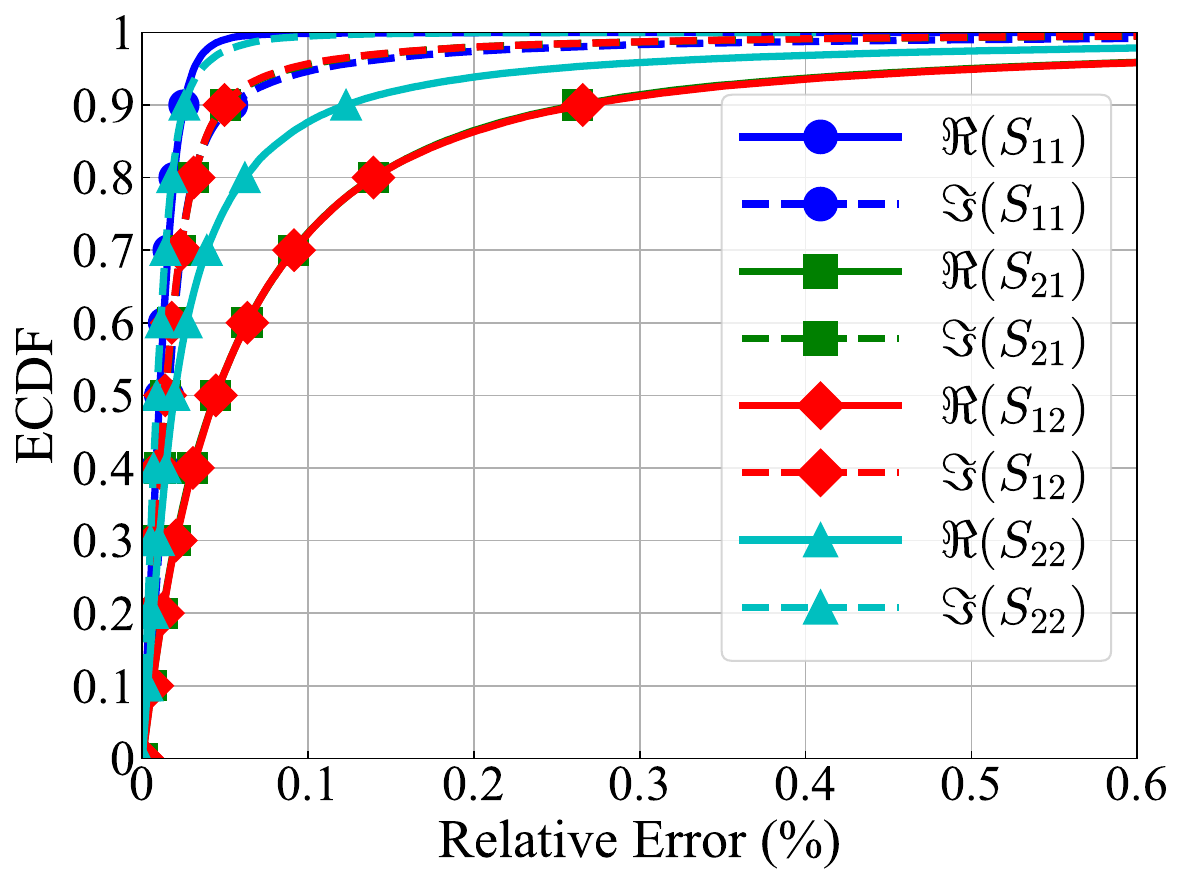}
    \label{fig:relative_error}
  }

  \caption{ECDF of the absolute and relative errors of RECBM-Net predictions for each real and imaginary component of the S-parameters on the test set.}
  \label{CDF_Aberror_Pererror}
\end{figure}

As shown in Fig. \ref{loss_curve}, the training loss converges to \(1.5 \times 10^{-8}\), while the validation loss stabilizes at \(9.1 \times 10^{-9}\). The extremely low validation loss provides initial evidence of the model's generalization ability. To further evaluate the RECBM-Net's ability to predict S-parameters, we constructed a test set of 100,000 samples, enabling a more comprehensive assessment of the model's performance on unseen data. The generation of this test set using Python takes 3.42 seconds. For each test sample, the frequency is uniformly sampled between 1.5 and 2 GHz, and the capacitor values for \(C_{4}\) and \(C_{5}\) are uniformly sampled between 0 and 10 pF. Fig. \ref{true_pred} presents a scatter plot comparing the predicted values and the ground truth S-parameters for 50 randomly selected samples from the test set. It can be observed that RECBM-Net achieves accurate predictions of the S-parameters. In addition, Table \ref{tab:mean_error} illustrates that the mean absolute error (MAE) of the predicted S-parameters in each dimension on the entire test set is on the order of \(10^{-5}\), while the corresponding mean relative error (MRE) is on the order of \(10^{-4}\), i.e., far below 1\(\%\). Averaging over all predictions, the overall MAE of the S-parameters is \(6.98 \times 10^{-5}\), and the overall MRE is 0.123\(\%\). More specifically, as depicted in Fig. \ref{CDF_Aberror_Pererror}, the error distribution computed over the entire test set reveals that in over \(95\%\) of the samples, the absolute error in each dimension of the predicted S-parameters is below \(3 \times 10^{-4}\), while the relative error is below \(0.6\%\). These results on mean error and statistical error distribution over the extensive test set collectively demonstrate that the model can predict the circuit behavior of a practical L-network with exceptional precision, thereby supporting the subsequent accurate determination of matching solution. 
\begin{figure}
\centering
\includegraphics[width=0.4\linewidth]{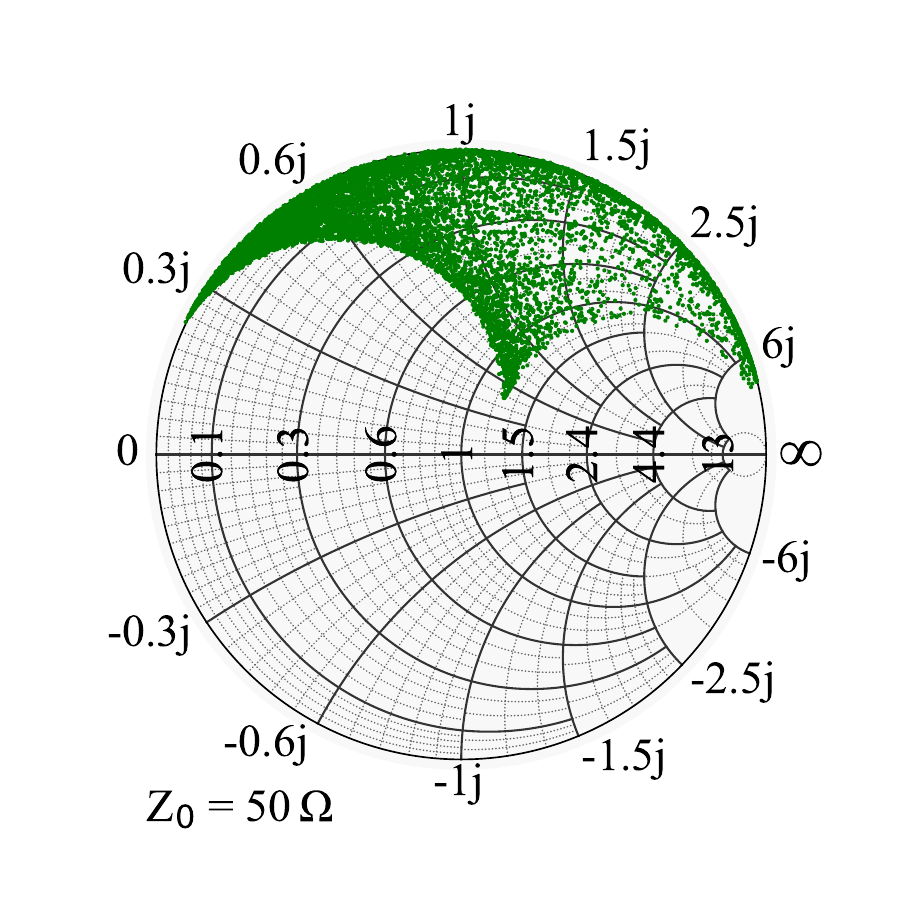}
\caption{Distribution of mismatched impedances under 9000 simulated mismatched scenarios over the frequency range from 1.5 GHz to 2 GHz.}\label{mis_imp} 
\end{figure}
\subsection{Performance of Data-Driven Impedance Matching Method} \label{section_5_2}
We simulate real-world mismatch conditions to verify the proposed data-driven impedance matching method and compare the performance of three matching solution determination strategies. We randomly generate 9000 combinations of \(f\), \(C_{4}^{*}\) and \(C_{5}^{*}\) within the tunable range. Assuming the input is perfectly matched (i.e., \(\Gamma_{in}= 0\)), the mismatched load reflection coefficient \(\Gamma_{L}\) is derived based on the practical L-network topology depicted in Fig. \ref{practical_Lnetwork}. This yields 9000 sets of load impedance and corresponding TMN configurations \((C_{4}^{*}, C_{5}^{*})\) that achieve perfect impedance matching. To simulate the mismatch, new values \((C_{4}^{\text{now}}, C_{5}^{\text{now}})\) for tunable capacitors are generated within the tunable range and the corresponding \(\Gamma_{\text{in}}\) is calculated based on the practical L-network circuit model. Finally, we generate a mismatched dataset that simulates real-world impedance mismatch. Each sample comprises the operating frequency \(f\), the TMN circuit configuration \((C_{4}^{\text{now}}, C_{5}^{\text{now}})\), the measured input port reflection coefficient \(\Gamma_{\text{in}}\), and the actual load reflection coefficient \(\Gamma_{L}\) corresponding to the current mismatched scenario. Since testing the adaptive impedance matching methods within the matching forbidden region is meaningless, this simulation method ensures that the mismatched impedance \(\Gamma_{L}\) remains outside the matching forbidden region.

\begin{figure}
\centering
\subfigure[ECDF of absolute error.]{\includegraphics[width=0.488\linewidth]{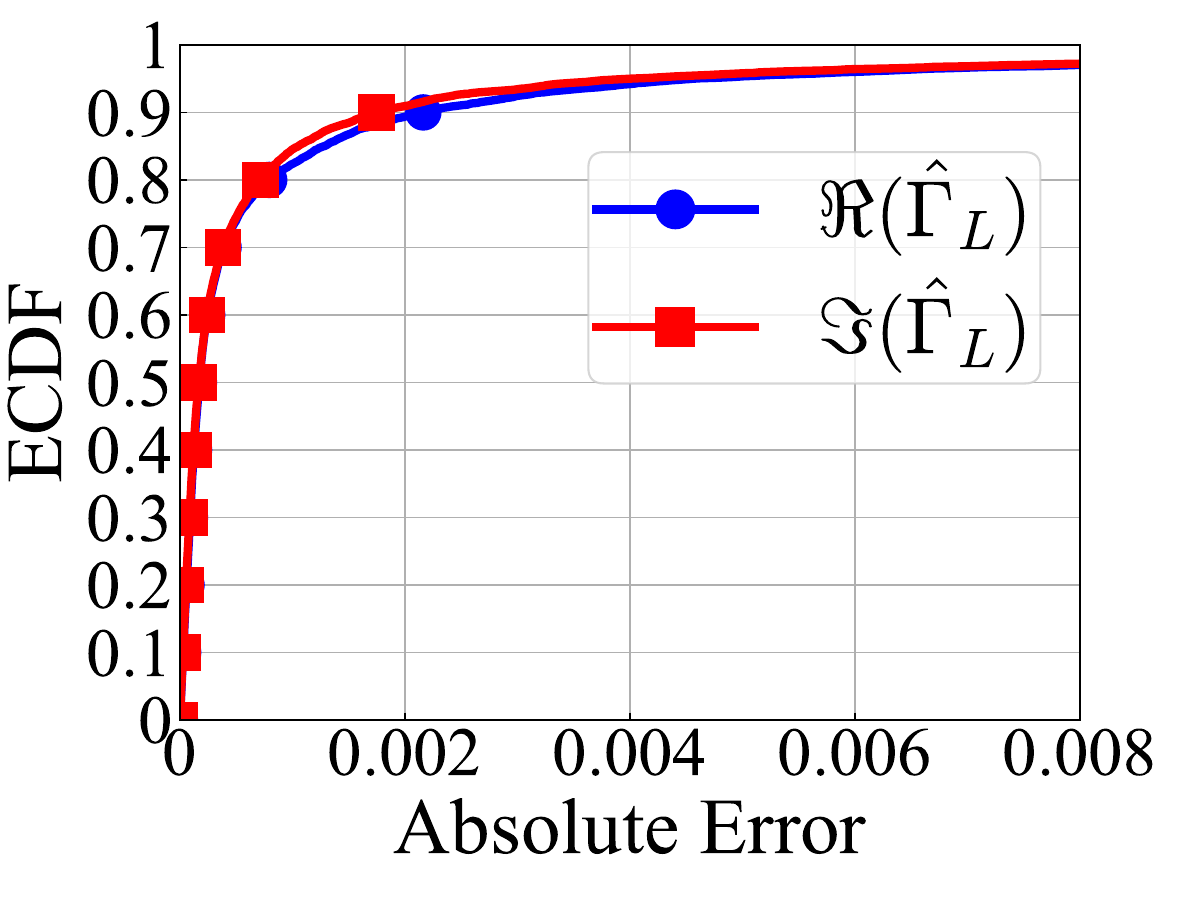}\label{absolute_error_load}}\hspace{0.1cm}
\subfigure[ECDF of relative error.]{\includegraphics[width=0.488\linewidth]{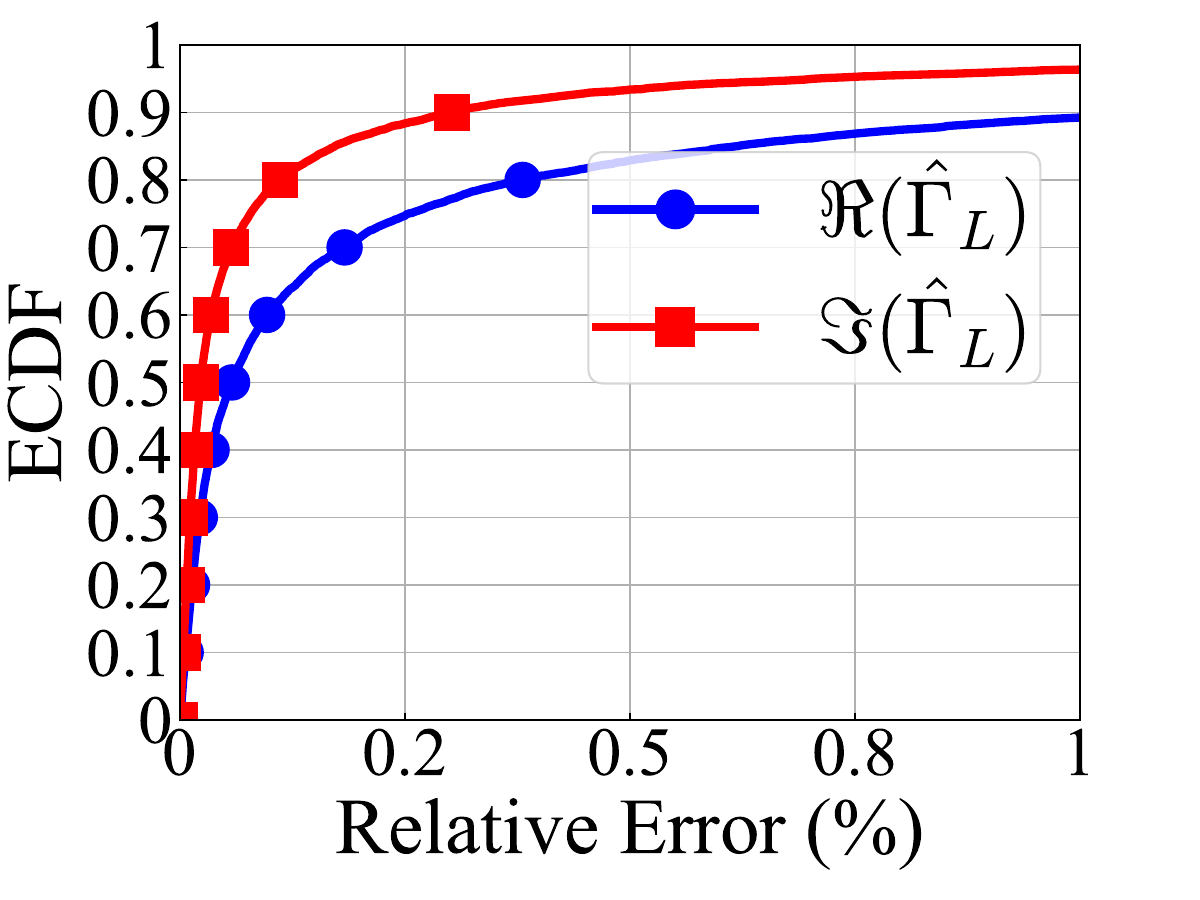}\label{percentage_errorr_load}}
\caption{ECDF of the absolute and relative errors between the true load reflection coefficient and that calculated from Eq. (\ref{eq:tao_L}) using the S-parameters predicted by RECBM-Net. The statistics are computed over 9000 simulated mismatched scenarios.}\label{CDF_Aberror_Pererror_load} 
\end{figure}

\begin{figure}
\centering
\subfigure[Predictions and ground truth.]{\includegraphics[scale=0.25]{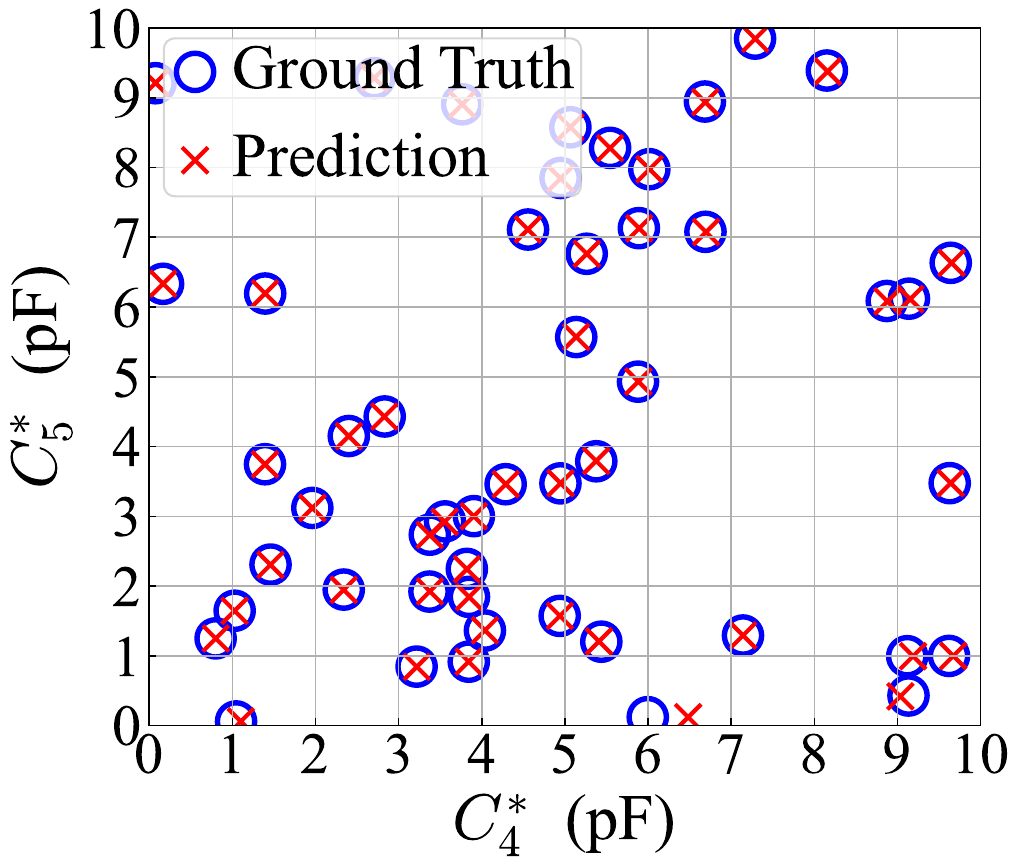}\label{error_solu}}\hspace{0.1cm}
\subfigure[ECDF of relative error.]{\includegraphics[scale=0.25]{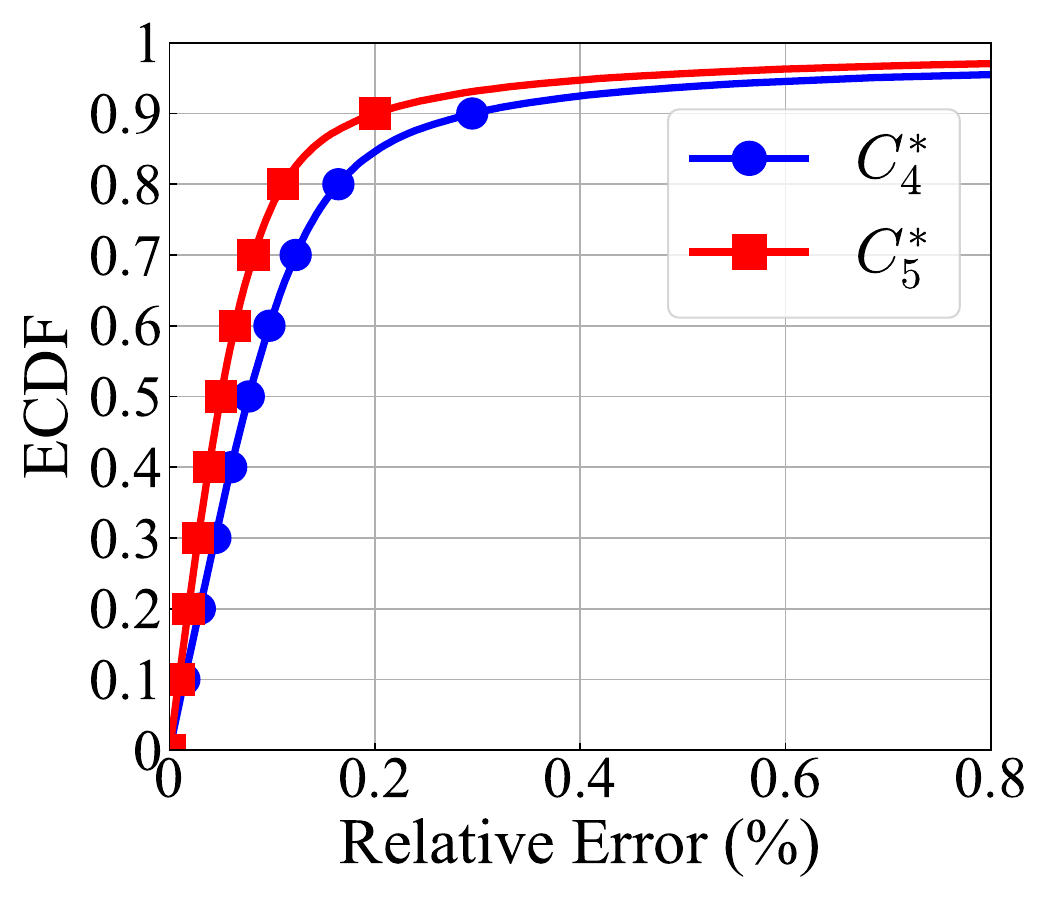}\label{percentage_errorr_solut}}
\caption{Comparison on the test set between the matching solution predicted by IMS-Net and the perfect matching solution derived from RECBM-Net.}\label{CDF_Aberror_Pererror_solution} 
\end{figure}

\begin{table}
\centering
\caption{Hyperparameter settings for the matching strategies SAPSO and AD-Adam}
\label{tab:hyperparameters}
\rowcolors{2}{white}{blue!10} 
\begin{tabular}{lcc}
\toprule
\textbf{Parameter} & \textbf{SAPSO} & \textbf{AD-Adam} \\
\midrule
Number of particles \(N\) & 50 & — \\
Individual learning factor \(\kappa_1\)& \(2.05\) & — \\
Social learning factor \(\kappa_2\)& \(2.05\) & — \\
Cooling factor \(\lambda\)& \(0.5\) & — \\
Initial solution \(\boldsymbol{\theta}_0\) & — & \((5\,\mathrm{pF},\,5\,\mathrm{pF})\) \\
Learning rate \(\alpha\)& — & 0.013 \\
Exponential decay rates \(\beta_1, \beta_2\)& — & \(0.9,\, 0.999\) \\
Stability constant \(\epsilon\)& — & \(10^{-8}\) \\
Maximum iterations \(\mathcal{I}_{\max}\)& \(100\) & \(500\) \\
Early stopping threshold \(\varepsilon\)& \(0.005\) & \(0.005\) \\
\bottomrule
\end{tabular}
\end{table}

As shown in Fig. \ref{mis_imp}, the simulated mismatched load impedance deviates significantly from 50 \(\Omega\). Following the impedance matching process described in Section \ref{section_4_2}, we first use RECBM-Net to infer the S-parameters for \((f,C_{4}^{\text{now}},C_{5}^{\text{now}})\) and calculate the predicted load reflection coefficient \(\hat{\Gamma}_{L}\). Fig. \ref{CDF_Aberror_Pererror_load} illustrates the empirical cumulative distribution functions (ECDFs) of the absolute and relative errors between the predicted \(\hat{\Gamma}_{L}\) and the actual \(\Gamma_{L}\) for all mismatched samples. Over 95\(\%\) of the samples exhibit absolute errors below 0.006 for both the real and imaginary parts of \(\hat{\Gamma}_{L}\), while over 90\(\%\) show relative errors below \(1\%\) in both parts. In summary, the load impedance estimated via the RECBM-Net exhibits excellent consistency with the actual impedance, thereby enabling the matching strategies to effectively minimize the actual input reflection coefficient.

Utilizing the dataset generation method described in Section \ref{section_4_2}, we construct the training dataset for IMS-Net. The traversal granularity for combinations of the operating frequency and TMN configurations is identical to that employed in the RECBM-Net training dataset (Section \ref{section_5_1}). Similarly, we generate a supervised learning dataset formatted as \(\{f,\Re (\hat{\Gamma}_{L}), \Im (\hat{\Gamma}_{L}),C_{p}^{*},C_{s}^{*}\}\) containing 6,500,000 samples with the MSE serving as the loss function. To amplify the gradient signal during backpropagation, the dataset labels (in pF) are scaled by a factor of 10. Subsequently, the dataset is partitioned into \(80\%\) for training and \(20\%\) for validation. IMS-Net is then trained on the same hardware platform using the Adam optimizer with an initial learning rate of 0.00002. Training is conducted for 3000 epochs with a batch size of 512, ultimately converging to training and validation losses of 0.052 and 0.051, respectively. To further validate IMS-Net's ability to predict matching solutions on unseen data, we generate a test set comprising 100,000 combinations by randomly sampling both the operating frequency and TMN configuration within a tunable range. Fig. \ref{CDF_Aberror_Pererror_solution} illustrates that IMS-Net reliably predicts matching solutions on the test set, with nearly \(95\%\) of samples achieving a relative error below \(0.6\%\). Notably, when \(C_{5}^{*}\) is extremely close to 0, the prediction accuracy exhibits a minor decline, while the overall performance remains robust.

Finally, we evaluate the end-to-end performance of the three matching solution determination strategies on the same 9000 mismatched samples generated earlier. For a fair comparison and improved computational efficiency, an early stopping threshold \(\varepsilon\) of 0.005 is applied to both methods. Empirical observations confirm that this threshold is reliably attainable and corresponds to excellent matching performance. For SAPSO, following the parameter settings in \cite{li2021simulated}, both the individual learning factor \(\kappa_{1}\) and the social learning factor \(\kappa_{2}\) are set to 2.05, while the cooling factor \(\lambda\) is set to 0.5. The number of particles \(N\) and the maximum iterations \(\mathcal{I}_{\max}\) are determined through hyperparameter tuning. For AD-Adam, the initial solution is set to the midpoint of the tunable capacitance range to reduce the number of required iterations. The exponential decay rates \(\beta_{1}=0.9\), \(\beta_{2}=0.999\), together with the stability constant \(\epsilon=10^{-8}\) are set according to the configuration in \cite{kingma2014adam}. The learning rate \(\alpha\) and the maximum iterations \(\mathcal{I}_{\max}\) are also determined through hyperparameter tuning. The detailed hyperparameter tuning procedures for both SAPSO and AD-Adam are provided in Appendix B of Supplementary Material. Table \ref{tab:hyperparameters} summarizes the hyperparameter settings for the two numerical optimization strategies. 

\begin{figure}
\centering
\subfigure[ECDF of median.]{\includegraphics[width=0.48\linewidth]{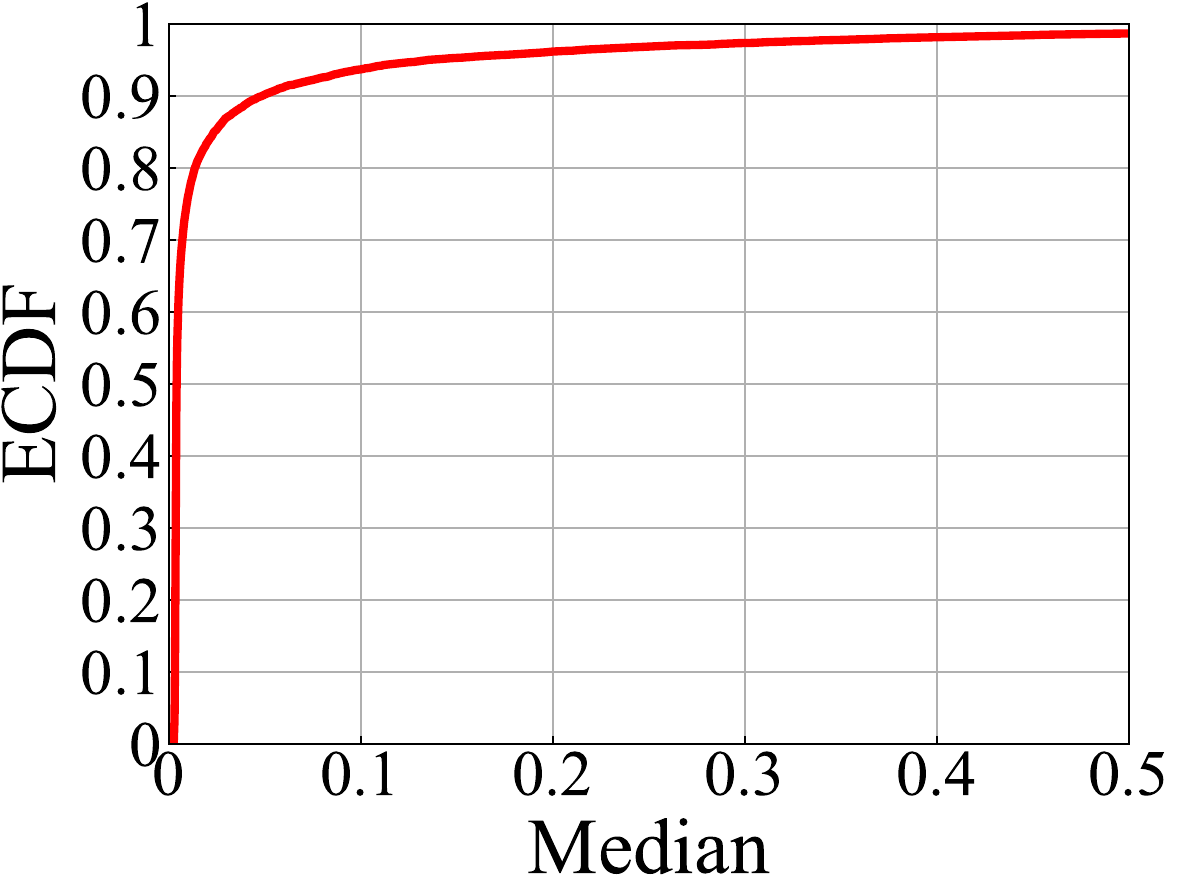}\label{SAPSO_median_cdf}}\hspace{0.1cm}
\subfigure[ECDF of SD.]{\includegraphics[width=0.48\linewidth]{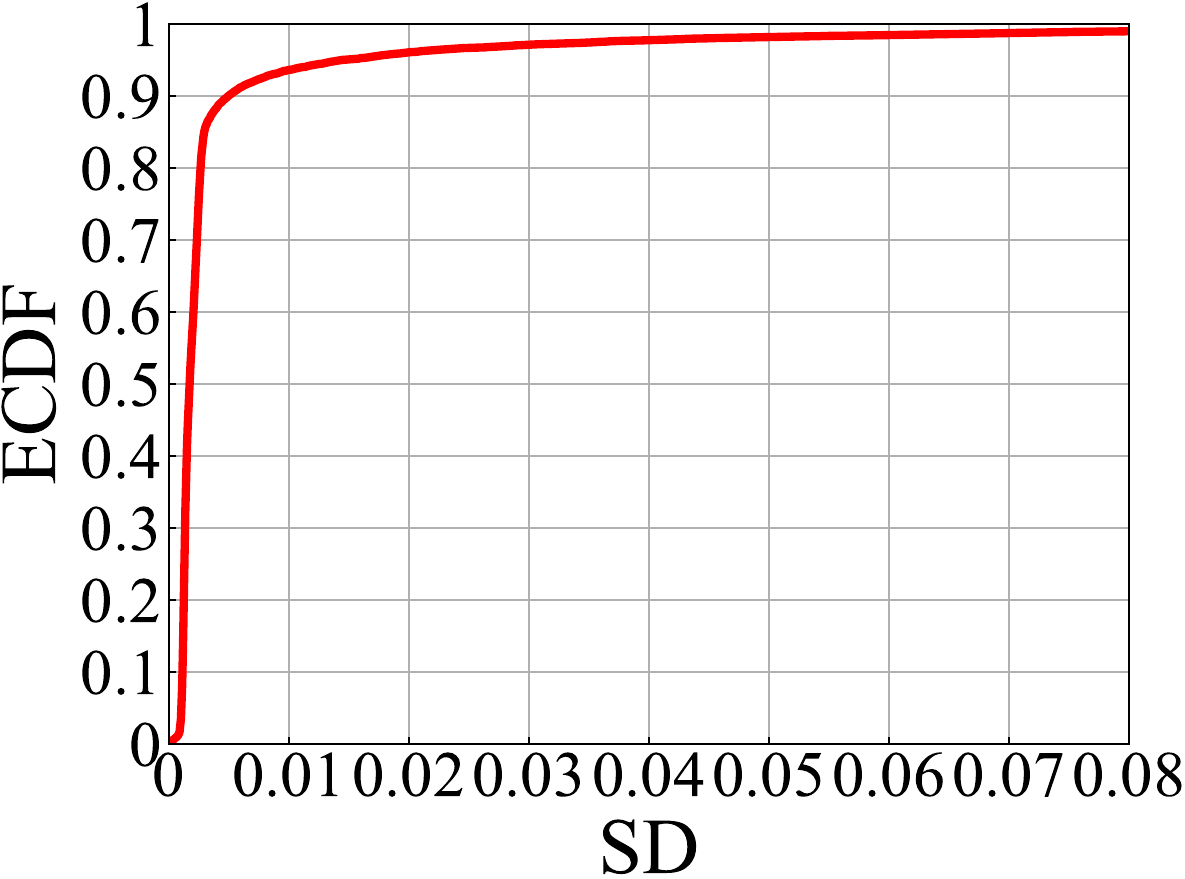}\label{SAPSO_std_cdf}}
\caption{ECDFs of the median and standard deviation of the reflection coefficient magnitude obtained by SAPSO over 30 independent runs per mismatched scenario across 9000 scenarios.}\label{CDF_SAPSO_medain_std} 
\end{figure}

For the stochastic SAPSO method, the mean, median, and standard deviation (SD) of the reflection coefficient magnitudes are calculated from 30 independent runs for each mismatched scenario. As shown in Fig. \ref{CDF_SAPSO_medain_std}, more than 95\(\%\) of the samples exhibit a median value below 0.1 and a SD below 0.01 across repeated runs, indicating excellent matching accuracy and stability. To evaluate the matching performance of SAPSO in each scenario, we use the mean tuned reflection coefficient magnitude obtained from multiple runs. In contrast, for other deterministic methods, the tuned reflection coefficient magnitude from a single run is used. 

\begin{figure}
\centering
\includegraphics[width=0.64\linewidth]{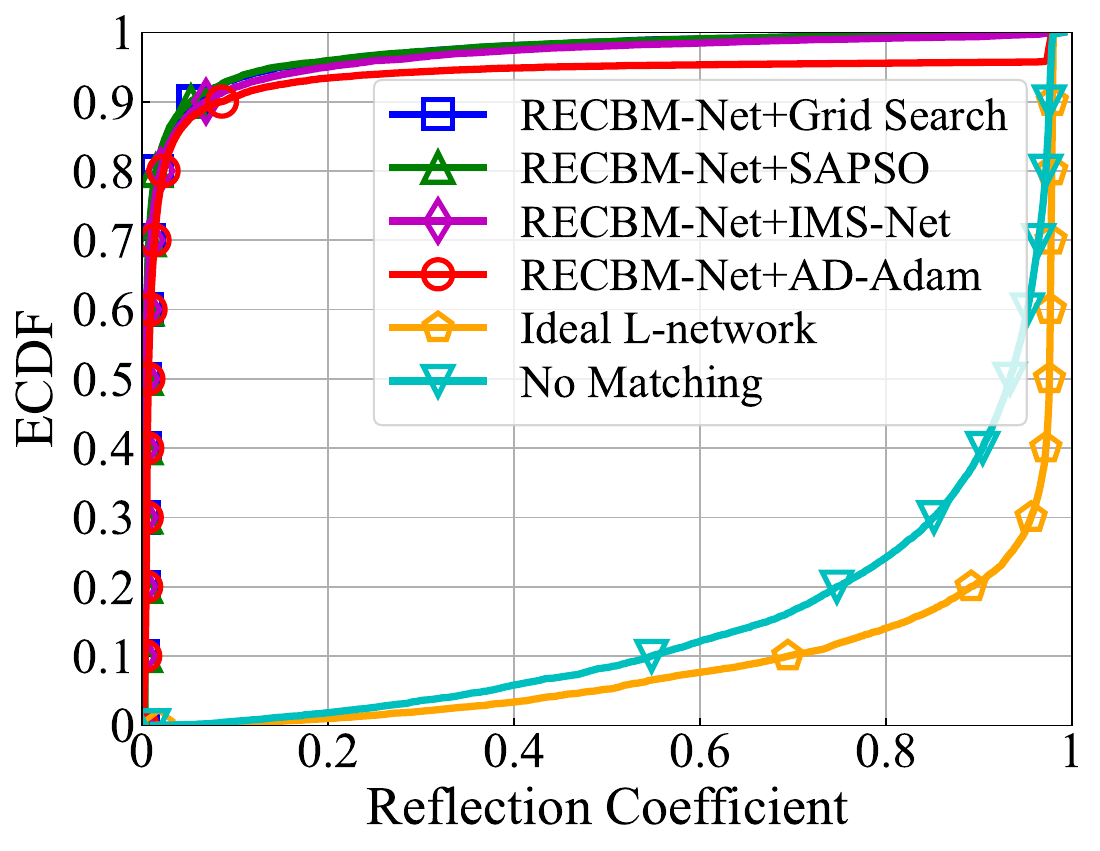}
\caption{Comparison of the matching performance of solutions obtained by the proposed data-driven method with different determination strategies, those analytically derived from the ideal L-network circuit model.}\label{Matching Reflection} 
\end{figure}

Fig. \ref{Matching Reflection} presents the ECDFs of the tuned reflection coefficient magnitudes obtained with different matching strategies across all test scenarios. It can been seen that when the load reflection coefficient is derived from the ideal L-network topology and the matching solution is computed using the closed-form expressions in Eq. (\ref{eq:Csolve}), the resulting solution even degrades the matching performance. In practical engineering applications, a reflection coefficient magnitude below 0.2 is generally considered a high-quality impedance matching \cite{kuchikulla2020applying}, corresponding to approximately 96\(\%\) of the incident power being delivered to the antenna. Based on this criterion, we define the compliance rate as the proportion of test samples with tuned reflection coefficient magnitudes below 0.2. The data-driven impedance matching method based on SAPSO achieves a compliance rate of 95.92\(\%\), slightly outperforming exhaustive grid search with a granularity of 0.01 pF \(\times\) 0.01 pF, which achieves 95.76\(\%\). IMS-Net also demonstrates high matching accuracy, achieving a compliance rate of 95\(\%\). In comparison, the AD-Adam method yields a lower compliance rate of 93.42\(\%\). This suboptimal performance can be attributed to the inherent limitations of gradient-based optimization algorithms when applied to non-convex objective landscapes.

In addition to the ECDFs of the tuned reflection coefficient magnitude for each matching strategy, we also summarize the overall mean, median, and SD of the tuned reflection coefficient magnitudes across all mismatched scenarios. As shown in Table \ref{tab:cross_statistics}, both SAPSO and IMS-Net achieve mean and median reflection coefficient magnitudes well below 0.2, with standard deviations around 0.1. These descriptive statistics demonstrate that both methods can consistently achieve low reflection coefficient magnitudes. Although AD-Adam exhibits a relatively larger SD, its mean and median remain well below 0.2. These results indicate that, despite its higher variability, the AD-Adam method is still capable of achieving effective impedance matching in most mismatched scenarios, even in the presence of parasitic effects.

\begin{figure}
\centering
\includegraphics[width=0.64\linewidth]{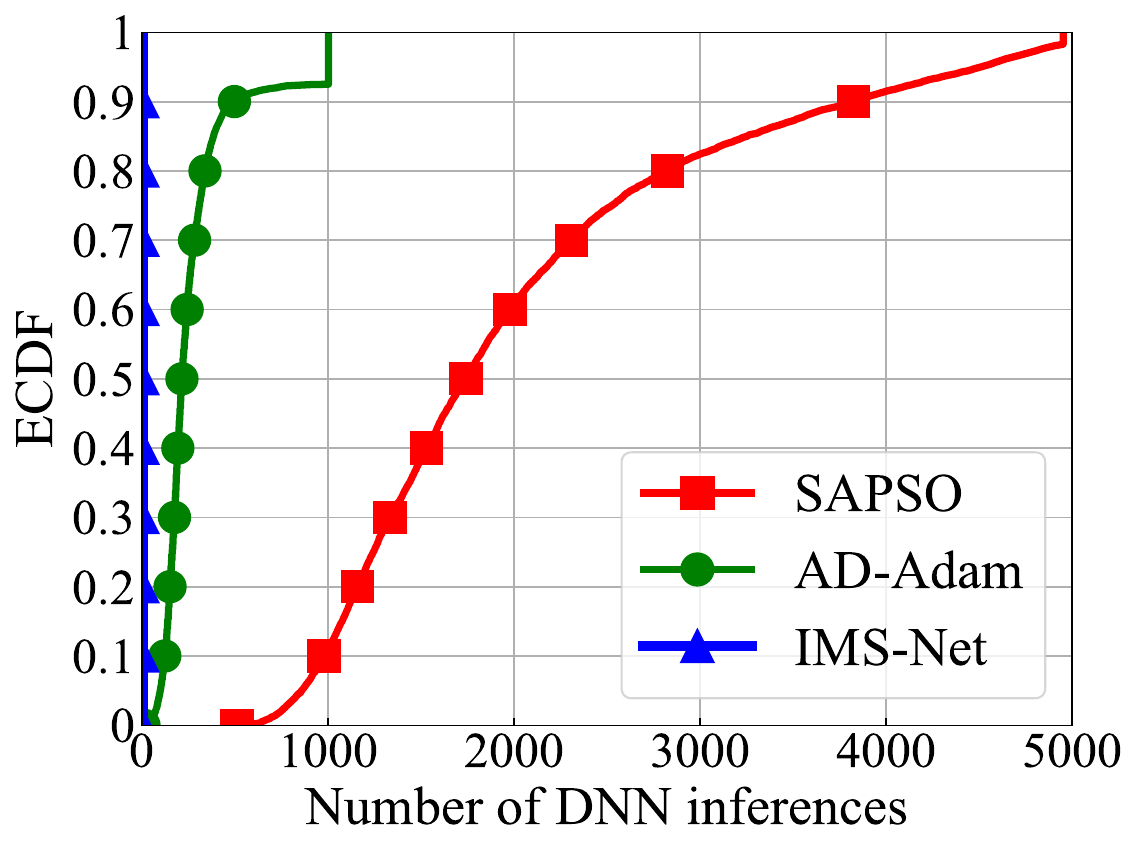}
\caption{ECDF of online computational overhead for three matching solution determination strategies (SAPSO, AD-Adam, and IMS-Net) across 9000 simulated mismatched scenarios. The computational overhead is quantified by the equivalent number of DNN inferences required.}\label{complex_compare} 
\end{figure}

Fig. \ref{complex_compare} presents the online computational overhead of the three matching solution determination strategies, evaluated over 9000 simulated mismatched scenarios. The overhead is quantified by the number of DNN inferences required, and their distributions are compared using ECDFs. Each strategy requires one RECBM-Net inference to compute the load reflection coefficient, which is included in the analysis. SAPSO achieves the highest matching accuracy among the three strategies, but incurs the highest computational overhead. Owing to its population-based search process, each sample requires an average of 2097 RECBM-Net inferences. For the AD-Adam method, the computational overhead of gradient evaluation in each iteration is approximately equivalent to that of a single RECBM-Net inference. Consequently, determining the matching solution with AD-Adam requires an average of 285 DNN inferences per sample. Furthermore, IMS-Net requires only one additional DNN inference for the matching solution, resulting in two inferences in total when including the inference for the load reflection coefficient. Although AD-Adam yields lower matching accuracy, it offers a substantial advantage in computational efficiency compared to SAPSO. IMS-Net achieves both high matching accuracy and minimal online computational overhead, but it requires an offline training, which introduces additional computational overhead prior to deployment. In summary, the proposed data-driven impedance matching method does not rely on explicit circuit topology and remains effective and accurate when the practical circuit topology deviates significantly from the ideal TMN.

\begin{table}
\centering
\caption{Descriptive statistics of the tuned reflection coefficient magnitudes for the three matching solution determination strategies}
\label{tab:cross_statistics}
\rowcolors{2}{white}{blue!10} 
\begin{tabular}{lccc}
\toprule
\textbf{Matching strategy} & \textbf{Mean} & \textbf{Median} & \textbf{SD}\\
\midrule
RECBM-Net+Grid Search  & 0.03062 & 0.00314 & 0.10214 \\
RECBM-Net+SAPSO & 0.03097 & 0.00403 & 0.10099 \\
RECBM-Net+AD-Adam& 0.06531 & 0.00648 & 0.20491\\
RECBM-Net+IMS-Net & 0.03717 & 0.00323 & 0.12107\\
Ideal L-network & 0.90261 & 0.97546 & 0.16649\\
\bottomrule
\end{tabular}
\end{table}
In addition to the performance analysis of the proposed data-driven impedance matching method, we also report the overall turnaround time for the algorithm development and testing using Python. First, RECBM-Net is trained to learn the circuit behavior. The entire process of dataset generation, model training, and testing takes approximately 59.45 hours. Next, to improve the performance of the numerical optimization methods, we derive a penalty function for SAPSO and perform hyperparameter tuning for both the SAPSO and AD-Adam methods. The corresponding development times are 39.54, 16.95, and 3.93 hours, respectively. Subsequently, IMS-Net is developed to directly predict matching solutions, where dataset generation, model training, and testing collectively require approximately 20.05 hours. Finally, to evaluate the end-to-end performance of the three matching solution determination strategies, we simulate 9000 mismatched scenarios. Dataset generation requires only 0.1422 seconds, while the evaluation times for SAPSO, AD-Adam, and IMS-Net across the 9000 scenarios are approximately 106.5 hours, 2.37 hours, and 0.00144 seconds, respectively. The development process described above is implemented sequentially on a single workstation. It is worth noting that the SAPSO development involves repeated experiments, which could be parallelized across multiple workstations to reduce the overall runtime. In summary, the overall experimental turnaround time for the proposed data-driven impedance matching method is approximately 248.79 hours. This time can be further reduced through the use of additional computing resources, highlighting the method’s potential to shorten the time-to-market (TTM).

\subsection{Performance under imperfect measurements} 
In this section, we first examine the impact of parasitic effects in the DUT on reflection coefficient measurement approaches applicable to small and portable platforms. A six-port reflectometer \cite{staszek2021balanced} provides a simple yet accurate means of measuring the reflection coefficient. It consists of a passive six-port network and four power detectors. The reflection coefficient of the DUT is computed from the scalar power measurements at the four detector ports. Since the computation depends solely on measured power and requires no knowledge of the DUT's internal circuit parameters, the resulting reflection coefficient inherently captures the effects of DUT parasitics rather than being distorted by them. Similarly, the three-point measurement method \cite{qiao2005antenna, xiong2019novel} based on a uniform transmission line offers another low-cost and practical solution for reflection coefficient measurement. In this approach, the DUT’s reflection coefficient is computed from voltage magnitudes sampled at three distinct locations along the line. This computation is likewise independent of the DUT's internal characteristics, ensuring that the measured reflection coefficient is not biased by parasitic effects.

Although these practical reflection coefficient measurement approaches are not affected by parasitics in the DUT, measurement noise remains unavoidable. To evaluate the impact of imperfect measurements on the proposed method, circularly symmetric complex Gaussian noise \(w\sim\mathcal{CN}(0, \sigma^2)\) is added to the input reflection coefficient \(\Gamma_{\text{in}}\), which is assumed ideally measured. Fig. 16 of Supplementary Material illustrates the matching performance of the three matching solution determination strategies under imperfect measurement conditions. As shown, all three methods exhibit performance degradation as the noise level increases. Specifically, for the SAPSO method, when the standard deviation of the measurement noise increases from \(\sigma=0.0002\) to \(\sigma=0.0004\), the compliance rate drops from 95.92\(\%\) (ideal case) to 94.44\(\%\) and 92.44\(\%\), respectively. Similarly, the compliance rate of the AD-Adam method decreases from 93.42\(\%\) to 91.23\(\%\) and 90.64\(\%\), respectively. For the IMS-Net method, the compliance rate declines from 95.00\(\%\) to 93.44\(\%\) and 91.55\(\%\), respectively.

Imperfect measurement of impedance mismatch metrics is a common issue that leads to degraded performance in adaptive impedance matching methods. The primary focus of this paper, however, is on developing an efficient and accurate adaptive impedance matching method that remains effective in the presence of parasitic effects.

\subsection{Feasibility analysis of embedded platform deployment} \label{5_3}

To assess the feasibility of implementing the proposed deep learning-based impedance matching methods in the radio unit illustrated in Fig. \ref{system}, we analyze their computational resource requirements. Specifically, we measure the static memory usage during deployment, the theoretical floating-point operations (FLOPs) associated with the deep learning component of the optimization process, and the peak runtime memory usage of each method.

RECBM-Net and IMS-Net contain 1,396,296 and 1,395,906 parameters, respectively, each requiring approximately 5.58 megabytes (MB) of single-precision storage. The number of parameters involved in the solution update procedures of SAPSO and AD-Adam is negligible compared with those of the DNN models. Consequently, both SAPSO and AD-Adam implementations require approximately 5.58 MB of static memory, whereas the method that directly determines the matching solution using IMS-Net requires \(5.58 \times 2=11.16\) MB. Across all three methods, the primary computational bottleneck arises from deep learning operations. Therefore, we measure the theoretical computational overhead of a single inference for RECBM-Net and IMS-Net to be 2.7872 MFLOPs and 2.7864 MFLOPs, respectively. Across the tested mismatched samples, the SAPSO method requires an average of 2097 RECBM-Net inferences per sample. Accounting for gradient computation overhead, the AD-Adam method incurs an average computational overhead equivalent to 285 RECBM-Net inferences for each mismatched sample. By contrast, IMS-Net obtains the matching solution with only one inference from RECBM-Net and one from IMS-Net. Multiplying the number of inferences by the corresponding theoretical inference FLOPs yields estimated computational overheads of 5.8448 GFLOPs, 0.7943 GFLOPs, and 0.0055 GFLOPs for the SAPSO, AD-Adam, and IMS-Net methods, respectively. Finally, the measured peak runtime memory usages for continuously determining matching solutions across mismatched samples are 587.35 MB for SAPSO, 792.62 MB for AD-Adam, and 362.09 MB for IMS-Net. A summary of these results is provided in Table \ref{tab:compute_overhead}. Although these metrics are obtained on an Intel Xeon Gold 5218 CPU, they characterize the inherent computational resource demands of the methods and thus provide a basis for evaluating deployment feasibility on embedded systems.

Based on the computational resource requirements of the three methods, we conduct a theoretical evaluation of their implementation on two embedded platforms. The CPU-based Raspberry Pi 5 \cite{weaver2024gflopsw} is equipped with a 16-GB microSD card for storage, which is sufficient to accommodate the model storage requirement of each method. It also provides 8 GB of random-access memory (RAM), which can fully satisfy the runtime memory requirements, as the peak runtime memory usage is on the order of several hundred megabytes. Considering the computational capacity of the Raspberry Pi 5 and the theoretical FLOPs of each method, the estimated computational latencies are approximately 186.1--279.2 milliseconds (ms) for SAPSO, 25.3--37.9 ms for AD-Adam, and 0.175--0.262 ms for IMS-Net (accounting for a 2--3\(\times\) multiplier due to framework overhead). The same approach is applied to evaluate the implementation of each method on the GPU-accelerated NVIDIA Jetson Nano \cite{NVIDIA2019}. It supports 16 GB of microSD card storage and is equipped with 4 GB of RAM, both sufficient to satisfy the storage and runtime memory requirements of each method. Owing to its higher computational capacity, the estimated computational latencies are significantly reduced, with 49.5--74.3 ms for SAPSO, 6.7--10.1 ms for AD-Adam, and 0.046--0.069 ms for IMS-Net.

In summary, the proposed deep learning-based impedance matching methods can theoretically be deployed on two representative embedded platforms. On hardware with lower computational capability, the SAPSO method may fall short of meeting real-time requirements. In contrast, the AD-Adam and IMS-Net methods can achieve millisecond-level real-time matching. The IMS-Net method, in particular, exhibits negligible latency in determining the matching solution, making it well suited for impedance matching applications that demand strict real-time performance.

\begin{table}
\centering
\caption{Computational complexity metrics for the three methods.}
\label{tab:compute_overhead}
\rowcolors{2}{white}{blue!10} 
\begin{tabular}{lccc}
\toprule
\textbf{Metric} & \textbf{SAPSO} & \textbf{AD-Adam} & \textbf{IMS-Net} \\
\midrule
Static storage size (MB) & 5.58 & 5.58 & 11.16 \\
Theoretical FLOPs (GFLOPs) & 5.8448 & 0.7943 & 0.0055 \\
Peak memory usage (MB) & 587.35 & 792.62 & 362.09 \\
\bottomrule
\end{tabular}
\end{table}

\section{Conclusion}\label{section_6}
In this work, we have proposed a data-driven adaptive impedance matching method that is robust to parasitic effects. First, we have designed RECBM-Net, a DNN that maps the operating state of a practical L-network to its corresponding S-parameters. Then, we have formulated the impedance matching task via the trained surrogate model as a mathematical optimization problem, and introduced two numerical optimization strategies with different online computational overheads. Finally, to avoid repeated RECBM-Net inference during matching, we have proposed IMS-Net that directly predicts the optimal solution in a single forward pass. 

Simulation results have shown that RECBM-Net achieves exceptionally high prediction accuracy of the S-parameters, with a MAE of \(6.98 \times 10^{-5}\) and a MRE of \(0.123 \%\) across both the real and imaginary components of all S-parameter entries. For 9000 simulated mismatched scenarios, SAPSO has reduced the magnitude of the reflection coefficient below 0.2 for 95.92\% of the samples but has incured substantial online computational overhead, averaging 2097 RECBM-Net inferences per sample. By exploiting gradient information, AD-Adam has significantly reduced the computational overhead to 285 inferences per sample, with a slight reduction in compliance to 93.42\%. IMS-Net has achieved the best trade-off between accuracy and computational overhead, requiring only a single inference to determine the matching solution while maintaining excellent performance, with the reflection coefficient magnitude reduced below 0.2 for \(95\%\) of samples.

In summary, all three matching solution determination strategies are necessary. SAPSO and AD-Adam serve as the respective baselines for matching accuracy and online computational overhead in comparison with IMS-Net. While IMS-Net offers the optimal trade-off between matching accuracy and computational overhead, it requires additional model training and testing. Therefore, when TTM constraints are critical, SAPSO and AD-Adam are more advantageous than IMS-Net, as both can be deployed immediately after training RECBM-Net. Under such constraints, SAPSO is suitable for impedance matching tasks that require high accuracy but are subject to low real-time constraints. A representative example is the compensation for impedance variations introduced during the manufacturing process after antenna design. For applications with rapidly varying antenna impedances that impose high real-time demands on the impedance matching process, AD-Adam is preferable owing to its low computational overhead. Finally, if the product development cycle allows for additional offline model training and testing, IMS-Net is the best choice as it provides matching accuracy comparable to SAPSO while achieving the lowest online computational overhead.

\ifCLASSOPTIONcaptionsoff
  \newpage
\fi
\bibliographystyle{IEEEtran}
\bibliography{reference.bib}

\begin{IEEEbiography}[{\includegraphics[width=1in,height=1.25in,clip,keepaspectratio]{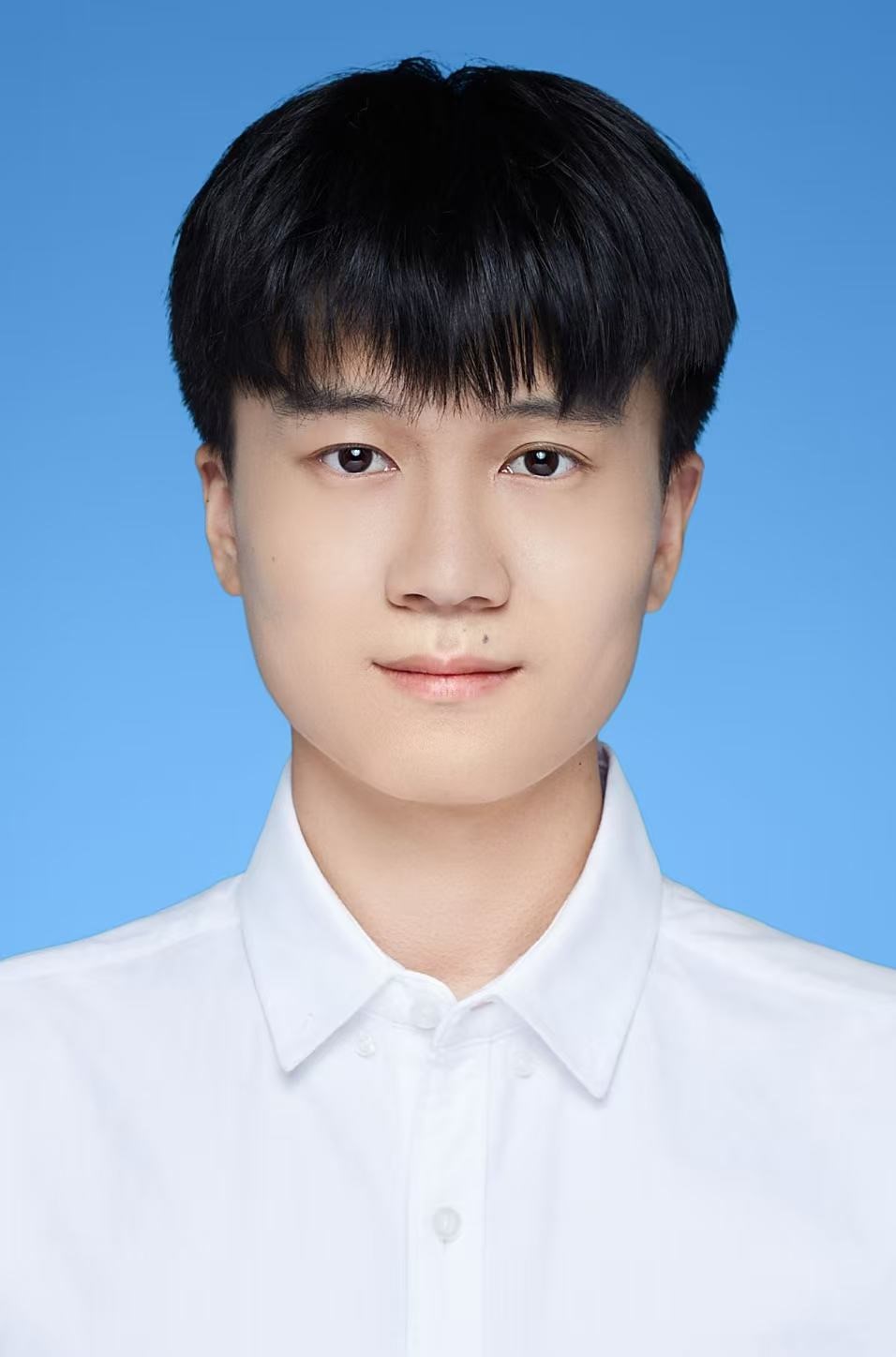}}]{Wendong Cheng}
    received his bachelor’s degree in communications engineering from the University of Electronic Science and Technology of China (UESTC), Chengdu, China, in 2023. He is currently pursuing the Ph.D. degree with the Department of Electronic Engineering and Information Science, University of Science and Technology of China (USTC), Hefei, China. His research interests include adaptive antenna impedance matching, deep learning for RF systems, and circuit theory.
\end{IEEEbiography}

\begin{IEEEbiography}[{\includegraphics[width=1in,height=1.25in,clip,keepaspectratio]{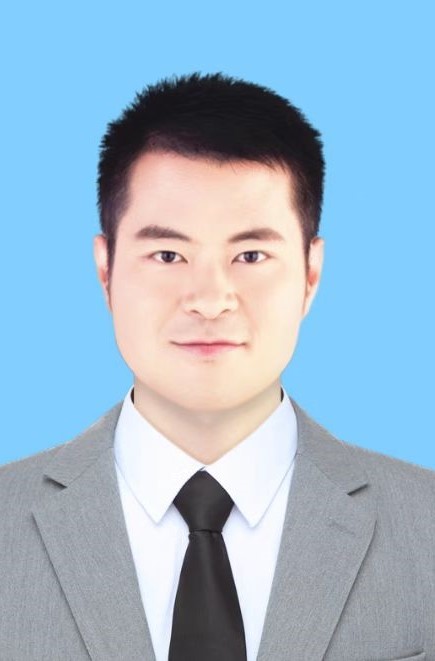}}]{Li Chen}
    (Senior Member, IEEE) received his B.E. degree in electrical and information engineering from the Harbin Institute of Technology (HIT), Harbin, China, in 2009, and his Ph.D. degree in electrical engineering from the University of Science and Technology of China (USTC), Hefei, China, in 2014. He is currently an Associate Professor at the Department of Electronic Engineering and Information Science, University of Science and Technology of China. His research interests include integrated computation and communication, and integrated sensing and communication.
\end{IEEEbiography}

\begin{IEEEbiography}[{\includegraphics[width=1in,height=1.25in,clip,keepaspectratio]{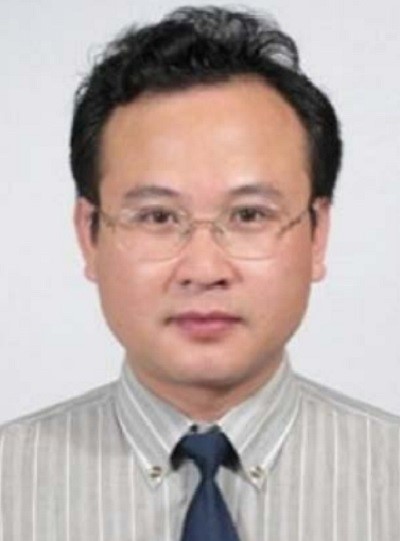}}]{Weidong Wang} 
    received the B.S. degree from Beijing University of Aeronautics and Astronautics, Beijing, China, in 1989 and the M.S. degree from the University of Science and Technology of China, Hefei, China, in 1993. He is currently a Full Professor with the Department of Electronic Engineering and Information Systems, University of Science and Technology of China. Prof. Wang is a member of the Committee of Optoelectronic Technology, Chinese Society of Astronautics. His research interests include wireless communication, microwave and millimeter-wave, and radar technology.
\end{IEEEbiography}

\clearpage
\includepdf[pages=-]{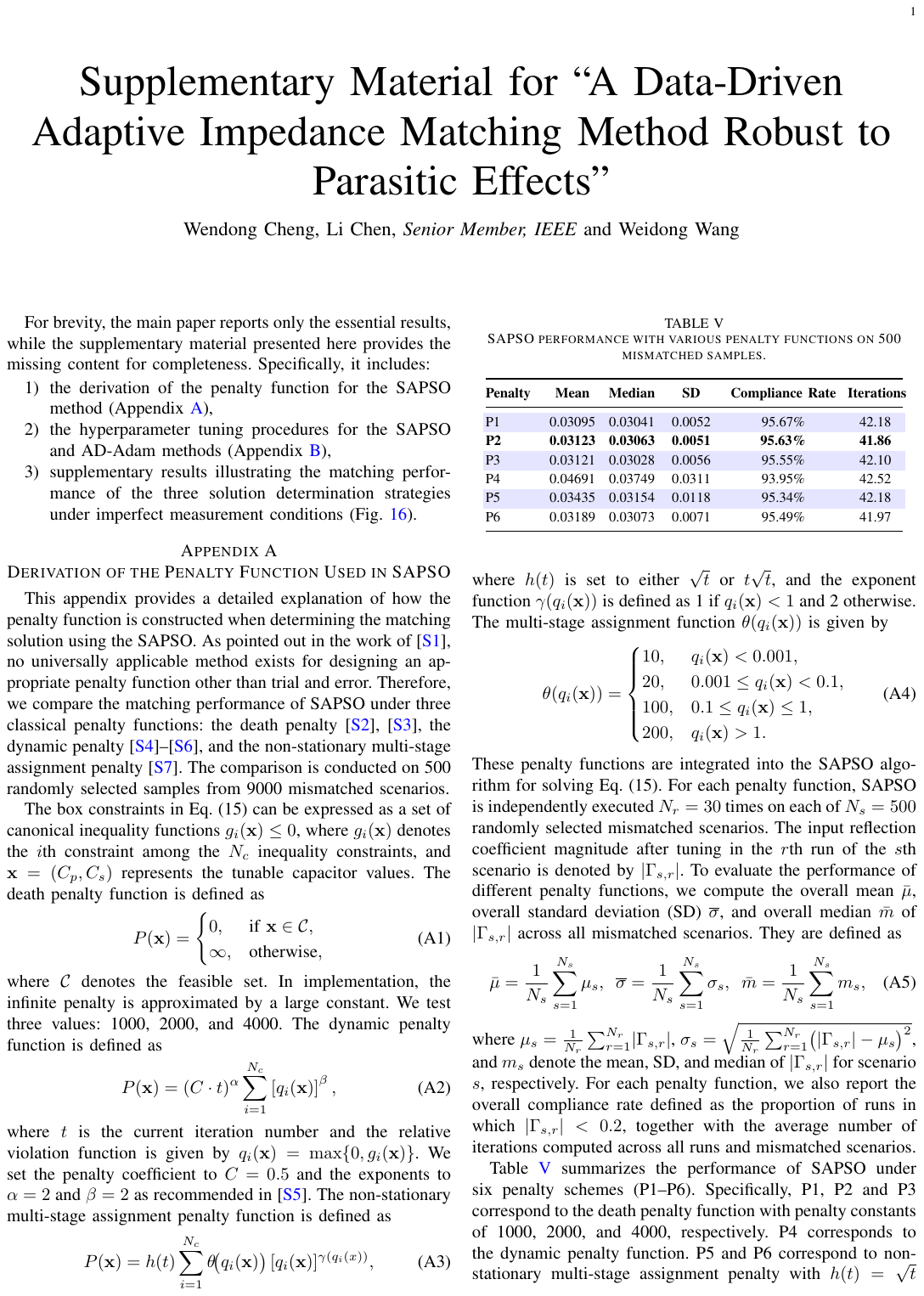}

\end{document}